\pdfoutput=1
\documentclass[11pt]{arxiv-article}

\bibliography{References}

\usepackage{Definitions}

\usepackage[utf8]{inputenc}

\RequirePackage{hyperref}

\newcommand{\OfficialTitle}{
 Matrix models at large charge
}
\title{\vspace{2cm}
  {\color{Thoughtless}\Huge\textbf{\dosserif\OfficialTitle}}
}

\hypersetup{pdfauthor={Orestis Loukas and Domenico Orlando and Susanne Reffert},pdftitle={\OfficialTitle}}

\author{%
  \begin{minipage}{.8\linewidth}
    \vspace{1cm}
    \begin{center} \dosserif
      {\small
         \textbf{Orestis Loukas},
         \textbf{Domenico~Orlando} and
        \textbf{Susanne~Reffert}}
    \end{center}
    \vspace{1cm}
    \authorBlock{}{Albert Einstein Center for Fundamental Physics\\
      Institute for Theoretical Physics\\
      University of Bern,\\
      Sidlerstrasse 5, \textsc{ch}-3012 Bern, Switzerland}
  \end{minipage}
}

\date{}

\begin{document}

\setstretch{1.15}

\numberwithin{equation}{section}

\begin{titlepage}

  \newgeometry{top=23.1mm,bottom=46.1mm,left=34.6mm,right=34.6mm}

  \maketitle

  \thispagestyle{empty}

  \vfill\dosserif

  \abstract{\normalfont \noindent

    We show that the large-charge formalism can be successfully applied to models that go beyond the vector models discussed so far in the literature.
We study the explicit example of a conformal $SU(3)$ matrix model in 2+1 space-time dimensions at fixed charge and calculate the anomalous dimension and fusion coefficients at leading order in the $U(1)$ charge.
  }

\vfill

\end{titlepage}

\restoregeometry

\tableofcontents

\section{Introduction}
\label{sec:introduction}

The Wilsonian effective action is very compelling from a conceptual point of view, but the fact that an infinity of terms compatible with the symmetries of the model in question appear in it drastically reduces its practical uses. There are however special sectors in which the Wilsonian action can live up to its conceptual power. In a sector of large fixed global charge $Q$, higher terms are suppressed by inverse powers of the large charge $Q$. Essentially, the Wilsonian effective action in a sector of large charge contains therefore only very few terms that are not suppressed and lends itself to explicit calculations of \ac{cft} data, such as the anomalous dimension and three-point functions.

\bigskip

In~\cite{Hellerman:2015nra}, field theories with global symmetries were studied in the sector where the value of the global charge $Q$ is large. It was shown subsequently~\cite{Alvarez-Gaume:2016vff} that the low-energy excitations of this sector are described by the general form of Goldstone’s theorem in the non-relativistic regime and that the effective field theory describing a sector of fixed $Q$ contains terms which are suppressed by inverse powers of $Q$. These results can be verified on the lattice and are in excellent agreement with the lattice computations~\cite{lattice}. 
Most of the existing literature has verified and extended the large-charge methods of~\cite{Hellerman:2015nra} for vector models of the $O(N)$ family~\cite{Loukas:2016ckj, Monin:2016jmo,Hellerman:2017efx}.\footnote{A notable and recent exception being~\cite{Hellerman:2017veg}.}
In this article, we venture to establish the applicability of large-charge approach beyond the class of vector models. The next logical step is to study models in 2+1 space-time dimensions at the \ac{ir} fixed point where the order parameter is a Hermitian traceless matrix, \emph{i.e.} lives in $A_{N-1}$ and the system has $SU(N)$ global symmetry.

\bigskip

$SU(N)$ matrix models are of interest as they are related to the  \(\CP{N-1}\) model which has been extensively studied in the condensed matter literature.
It is believed to flow to a conformal fixed point of the type we discuss here. Here, we will concentrate on the Noether charge and not consider the physics associated to possible topological symmetries.
In the \ac{uv}, the  \(\CP{N-1}\) describes a compact gauge field (associated to magnetic monopole defects) coupled to an $N$-component complex scalar that satisfies a norm constraint (see~\cite{Coleman:1988aos} for a pedagogical introduction).
At the critical point (which cannot be easily accessed starting from a \ac{lg} description), its universality class is believed to describe the quantum transition between an $SU(N)$ lattice antiferromagnet\footnote{Generally, $SU(N)$ antiferromagnetic Heisenberg spin systems with $N > 2$ serve  to model many physical systems ranging from spin-orbit coupled transition metal compounds~\cite{Kugel:1982}, to ultracold atoms in optical lattice potentials~\cite{2010NatPh.6.289G,Laflamme:2015wma}.} and a valence-bond-solid~\cite{PhysRevLett.71.1911,PhysRevB.70.075104}.
The connection between the phase transition in a microscopic Hamiltonian and a low-energy continuum theory description  still remains to be verified, though.
Therefore, in order to provide evidence for such a connection, it is of utmost importance to compare various universal properties arising from those two different descriptions. 

One of the reasons why using a standard \ac{lg} approach to reach the critical point is complicated is that the \(\CP{N-1}\) model is not invariant under parity.
There is in fact experimental evidence that the model undergoes a second-order phase transition for \(N = 2, 3\), a first-order phase transition for \(N \ge 4\), and again a second-order one for \(N \gg 1\)~\cite{2013PhRvB.88m4411N}.
On the analytic side, valuable results have come from a large-$N$ expansion of the $\CP{N-1}$ model~\cite{Murthy:1989ps,PhysRevB.42.4568}. More recently, the $1/N$ expansion in conjunction with the state-operator correspondence of conformal field theory was used to study magnetic monopole operators at the critical point of the $\CP{N-1}$ model. In \emph{e.g.}~\cite{PhysRevB.78.214418,Pufu:2013eda,Dyer:2015zha} the derived scaling dimension of monopole operators was compared with the power-law decay of the valence bond solid at the quantum critical point~\cite{2004Sci.303.1490S,PhysRevB.70.144407}.

A parity-invariant generalization %
has been proposed in~\cite{2015PhRvE.91e2109D} and in the special case of \(N = 3\) it is conjectured that the parity-invariant model exhibits a symmetry enhancement at the critical point which is in the same universality class as the $O(8)$ model. Evidence for that via lattice simulations is provided in~\cite{PhysRevB.85.180411}.

\bigskip
Our approach to studying the large-charge sector of $SU(N)$ matrix models is similar to the one used in~\cite{Alvarez-Gaume:2016vff}. We start by writing an effective Wilsonian action in 2+1 dimensions which must be at least approximately scale-invariant. We look for homogeneous, fixed-charge ground states and expand around the ground state to find the symmetry-breaking pattern. We show that, like in the case of the vector model, large charge suppresses all interactions.
Our approach is quite general and the results generalizable, but we have chosen to concentrate on the $N = 3$ case, as the algebra is much more tractable than for $N\geq 4$. In order to provide concrete results, we compute the conformal dimension and a three-point function for the $SU(3)$ matrix model. An interesting observation, and ultimately the reason for the simplicity of our final results, is that at leading order in the charge, the model exhibits an Abelian structure as the low-energy physics is governed by a single relativistic Goldstone boson. Despite starting from a global $SU(3)$ symmetry, the effective action at leading order resembles the ones of simpler Abelian cases and the explicit results are similar to the ones found in~\cite{Hellerman:2015nra,Alvarez-Gaume:2016vff,lattice,Loukas:2016ckj,Monin:2016jmo}.

\bigskip

Concretely, we start by writing a linear sigma model in the \ac{ir} to find the ground state and
 symmetry-breaking pattern.
Like in the vector model, we find that a homogeneous ground state is possible only if we fix a single $U(1)$.
The symmetry-breaking pattern however presents a surprise: when fixing a $U(1)$ charge one would expect the
     $SU(3)$ symmetry to be explicitly broken to $U(1)^2$. This is however not the case here: at leading order in the charge, there is an accidental symmetry enhancement and the explicit breaking
     is to $U(2)$, for any choice of the fixed $U(1)$ direction in the Cartan subalgebra. The spontaneous symmetry-breaking pattern is then $U(2) \to U(1)$ and there will be three
     Goldstone \acp{dof}.
Next, we write a non-linear sigma model for these Goldstone \ac{dof} and find that
 the situation is very similar to the one in the vector models,
 with one relativistic and one non-relativistic Goldstone field.
Finally, we use the \ac{ccwz} formalism (as suggested in~\cite{Monin:2016jmo})
 to compute the three-point function for the insertion of two operators of
 large charge and one of generic charge and we compute the
 large-charge behavior of the relevant fusion coefficients.

 \bigskip
 
Extending the large-charge approach to matrix models yields results which are expected based on our experience with the vector model~\cite{Alvarez-Gaume:2016vff}, testifying to the \emph{general applicability of our approach}. The matrix models however exhibit a richer behavior than their simpler cousins, giving rise to some phenomena that had not appeared before.
As in the case of the vector models, we find that if we want a homogeneous ground state, at least in $N\leq3$, we can only fix one $U(1$) charge (\emph{i.e.} a direction in the maximal torus).
The low-energy (large-charge) physics is fixed by the same symmetry-breaking pattern as in the vector models
and the large charge controls strong coupling.
There is a simple formula for the conformal dimensions which is essentially the same as in the vector models.
The matrix models however also show some new and unexpected behavior:
the effective potential depends on two parameters. For some
 values of these parameters, it is not possible to fix a generic $U(1)$
 charge.
We find moreover an accidental symmetry enhancement at large charge.

Our concrete results for the case of $SU(3)$ are generalizable, but not general. We can use the very same formalism to analyze any symmetry group $SU(N)$. For $N > 3$, it turns out that there are homogeneous solutions with more than one charge --- this does not happen in the vector models or in $SU(3)$; presumably the physics will be similar to the case of $O(2)\times O(2)$ discussed in~\cite{Monin:2016jmo}.%

\bigskip

The plan of this paper is as follows.
We start out with the linear sigma model description of the $SU(N)$ matrix models at large charge in Section~\refstring{sec:linear-sigma-model}, finding the symmetry breaking patterns associated to homogeneous solutions with one fixed charge.
In Section~\refstring{sec:non-linear-sigma-model} we discuss the non-linear sigma model realization for $SU(N)$ matrix theory at large charge, which is more general than the linear sigma model description.
In Section~\refstring{sec:CCWZ} we use the \ac{ccwz} formalism to explicitly compute the fusion coefficients for the $SU(3)$ matrix model.
In Section~\refstring{sec:conclusions} we end with conclusions and outlook. In Appendix~\ref{app:Conventions} we collect the conventions for the algebra $A_2$.

\section{Linear sigma model}
\label{sec:linear-sigma-model}

In this section, we set up our problem in terms of a linear sigma model. We will make an ansatz for the effective action at the conformal point, assuming scale invariance and including terms with at most two derivatives. We will then look for homogeneous solutions and determine the associated symmetry breaking pattern.

The starting point of our analysis is the Wilsonian effective action for a theory with order parameter $\Phi \in A_{N-1}$. The Wilsonian action is often considered to be of limited use as it contains infinitely many higher operators. The large-charge limit has however the power to turn it into a useful object. In particular, we will see that we do not have to \emph{postulate} a symmetry-breaking pattern, but are able to derive it from the analysis of the action.

Conformal symmetry requires scale invariance of the action and fixes the potential to be a polynomial of order six in $\Phi$ and the conformal coupling of $\Phi$.
For the time being, we neglect higher-derivative operators.
We will show in the following that these contributions are controlled in the large-charge expansion.
Given these assumptions, an effective \ac{ir} Wilsonian action for $\Phi$ living in $\setR\times\Sigma$ (where $\Sigma$ is a two-dimensional surface) is given by
\begin{equation}\label{eq:L}
  S= \int_{\setR\times
    \Sigma} \dd{t} \dd{\Sigma} \mathcal{L} =\int_{\setR\times
    \Sigma} \dd{t} \dd{\Sigma} \bqty{\frac{1}{2} \Tr (\del_\mu \Phi \del^\mu \Phi) - \frac{R}{16} \Tr\Phi^2 - V(\Phi)},
\end{equation}
where $R$ is the scalar curvature of $\Sigma$ and
\begin{equation}
  V(\Phi) = g_1 \Tr \Phi^6 + g_2 (\Tr\Phi^3)^2  + g_3 \Tr\Phi^4\Tr\Phi^2 + g_4 (\Tr\Phi^2)^3.
\end{equation}
We will use this action to find the symmetry-breaking pattern associated to fixing the charge.
The Euler--Lagrange \ac{eom} are found by varying $S$ with respect to $\Phi$:
\begin{equation}
\label{LSM:EulerLagrangeEOMs}
        \ddot \Phi = - V'(\Phi).
\end{equation}

For the purpose of this work, we will limit ourselves to the case of $N=3$, \emph{i.e.} $\Phi \in A_2$, where the action is invariant under the adjoint action of the group $SU(3)$. This simplifies the form of the potential due to the identities
\begin{align}
  \Tr\Phi^4 &= \frac{1}{2} \left( \Tr\Phi^2\right)^2, &
                                                           \Tr\Phi^6 &= \frac{1}{3} \left(\Tr\Phi^3\right)^2 +  \frac{1}{4} \left(\Tr\Phi^2\right)^3,
\end{align}
which are ultimately a consequence of the fact that there are only two invariant symmetric tensors in $A_2$, namely the identity and the $\mathrm{d}$-tensor.
In fact, decomposing $\Phi$ on an appropriate basis of generators $\lambda^a=2T^a$ of the algebra as $\Phi=\phi_a \lambda^a$, we find using Eq.~\eqref{app:Product_GellMannMatrices},
\begin{align}
  \Tr\Phi^2 &=%
              2\delta_{ab}\phi^a\phi^b, \\
  \Tr\Phi^3 &=%
              \mathrm{d}_{abc}\phi^a\phi^b\phi^c,
  \\
  \Tr\Phi^6&=
\frac13d_{abc}d_{a'b'c'}\phi^a\phi^b\phi^c \phi^{a'}\phi^{b'}\phi^{c'}
+
2\delta_{ab}\delta_{a'b'}\delta_{a''b''}\phi^a\phi^b\phi^{a'}\phi^{b'}\phi^{a''}\phi^{b''}
.
\end{align}
This means that we can choose any two order-six polynomials in $V$ to write the most general potential, \emph{e.g.}
\begin{equation}
  V(\Phi) = g_1 \Tr\Phi^6 + g_2 \pqty{\Tr\Phi^2}^3.
\end{equation}
In the special case of $g_1=0$, the symmetry of the model is enhanced to $O(8)$, acting on the vector with the components $(\phi^1,\dots,\phi^8)$, bringing us back to the vector model studied in~\cite{Alvarez-Gaume:2016vff}.

The model is only consistent if \(V(\Phi)\) is bounded from below.
Since $V$ is a function of $\Tr\Phi^n$, it is enough to consider the eigenvalues \(\set{a_1, a_2, -(a_1 + a_2)}\) of \(\Phi\).
The potential is bounded from below if it goes to \(+ \infty\) when \(a_1\) or \(a_2\) diverge.
If we introduce the combinations
\begin{align}
  \label{eq:alternative-couplings}
  g_0 &= \frac{1}{4} g_1 + g_2 , & \delta &= \frac{11 g_1 + 36 g_2}{g_1 + 4 g_2},
\end{align}
we find that the boundedness is assured if both $\lambda$ and $\delta$ are strictly positive:
\begin{align}
  g_0 &>0, & \delta &>0 .
\end{align}
This is only a necessary condition, though.
We will see in the following that general homogeneous fixed-charge solutions may require more stringent conditions on the parameters.

\subsection{Homogeneous ground state}
\label{sec:homogeneous-ground-state}

In the spirit of~\cite{Alvarez-Gaume:2016vff,Hellerman:2015nra}, we look for the most general \emph{homogeneous} solutions to the \ac{eom} stemming from the Wilsonian effective action.\footnote{This is not to say that more general inhomogeneous configurations do not exist or are not interesting, see~\cite{Hellerman:2017efx} for an analysis of inhomogeneous solutions at fixed charge in the $O(4)$ model.}
If the system is compactified on $\setR\times S^2$, the state-operator correspondence will map the quantum state to a scalar primary inserted at the origin: the energy of the state coincides with the dimension of the operator.

The matrix \(\Phi\) is Hermitian, so we can diagonalize it as
\begin{equation}
\Phi = UAU^\dagger,
\end{equation}
where $U$ is unitary and $A$ is a real traceless diagonal matrix:
\begin{align}
  A &= \pmqty{\dmat{a_1, a_2, \ddots, a_N}}, & a_1 + \cdots + a_N &= 0 .
\end{align}

The $SU(N)$ symmetry of the action is reflected in the existence of a conserved Noether current
\begin{equation}
  J_\mu =  i B\comm{\Phi}{\del_\mu \Phi} ,
\end{equation}
where $B$ is some diagonal matrix.
For $N=3$, $B=b\cdot\Id$, where $b$ is a real parameter chosen such that the conserved charge will be quantized independently of the global properties of $\Phi$ which are not fixed by the symmetries.
We will be mostly interested in the charge density \(J_0\), which can be rewritten as
\begin{equation}
  \label{eq:J0Current}
  J_0 = b U \comm{\comm{\omega}{A}}{A} U^\dagger ,
\end{equation}
where \(\omega\) is the angular velocity
\begin{equation}
  \omega = -i U^\dagger \dot U .
\end{equation}
The conserved charge $Q$ is defined as
\begin{equation}
\int_\Sigma J_0 = Q,
\end{equation}
which in the case of a homogeneous solution becomes $Q= J_0 \cdot \mathrm{Vol(\Sigma)}$, and hence $\dot J_0 =0$.
It is convenient to introduce also the matrix
\begin{equation}
  K =  U^\dagger J_0 U
\end{equation}
and think of it as the momentum associated to $\omega $:
\begin{equation}
  K= b \pqty{ \pdv{\mathcal{L}}{\omega}}^T  = b \comm{\comm{\omega}{A}}{A} .
\end{equation}
This allows us to write the Hamiltonian corresponding to the Lagrangian given in Eq.~\eqref{eq:L} in a compact form:
\begin{equation}
  \label{eq:linear-sigma-Hamiltonian}
  \mathcal{H} = \frac{1}{2} \Tr( \pi_A^2 + (\nabla A)^2 + \comm{U^\dagger\nabla U}{A} + 2 V(A) ) + \frac{1}{2b^2} \sum_{i\neq j} \frac{\abs{K_{ij}}^2}{\pqty{a_i - a_j}^2},
\end{equation}
where $\pi_A^T = \delta \mathcal{L}/\delta \dot A$.
The relation between \(K\) and \(\omega\) expressed in components is
\begin{equation}
  K_{ij} = b  \omega_{ij} \pqty{a_i - a_j}^2,
\end{equation}
which implies that the diagonal components of $K$ vanish identically, $K_{ii} = 0$.

Since we are looking for homogeneous solutions ($\nabla \Phi=0$), we have an effective \ac{qm} problem, for which powerful methods originating from integrability have been developed.
We introduce the Lax matrix
\begin{equation}
\label{def:LaxMatrix}
  L = -i U^\dagger \dot \Phi U = \dot A+ i \comm{\omega}{A}
\end{equation}
in order to write the \ac{eom} \eqref{LSM:EulerLagrangeEOMs} as
\begin{equation}
  \label{eq:eom}
  \dot L + i \comm{\omega}{L} = - V'(A).
\end{equation}

\paragraph{The diagonal part of the EOM.}

Let us study first the diagonal part of this equation.
The \ac{lhs} lives by construction in the algebra, which consists of traceless matrices, so the \ac{eom} implies
\begin{equation}
  \Tr[V'(A)]=0.
\end{equation}
In our case,
\begin{equation}
\label{LSM:OriginalPotential}
  V(A) = \frac{R}{16} \Tr A^2 + \frac{1}{6} \left( g_1 \Tr A^6 + g_2 \Tr(A^2)^3\right).
\end{equation}
We can parametrize $A$ in terms of Gell--Mann matrices:\footnote{See Appendix~\ref{app:Conventions} for the conventions used here.}
\begin{equation}
  A = \pmqty{\dmat{a_1, a_2, a_3}} =
  \pmqty{\dmat{ \alpha_1 + \alpha_2, -\alpha_1+\alpha_2 ,-2 \alpha_2}} =
  \alpha_1 \lambda_3 + \alpha_2 \lambda_8\sqrt 3.
\end{equation}
Like this, the trace condition becomes
\begin{equation}
  \Tr[V'(A)] = \Tr \left[\frac{R}{8} A + g_1  A^5 + g_2 \Tr(A^2)^2A \right] = g_1 \alpha_2\left( (\alpha_1^2+\alpha_2^2)^2-4\alpha_2^4\right) = 0.
\end{equation}
For the case of $g_1 \neq 0$, where the model does \emph{not} reduce to the $O(8)$ vector model, this is only satisfied (up to trivial permutations) if $\alpha_2=0$, \emph{i.e.} if $A = \alpha_1 \lambda_3$. It follows that
\begin{equation}
  V'(A) = \left(\frac{R}{8}a_1 +4g_0a_1^5\right) \begin{pmatrix}1 \\ & -1 \\ &  & 0 \end{pmatrix} = \left(\frac{R}{8}a_1 +4g_0a_1^5\right) \lambda_3 .
\end{equation}

It is convenient to write the matrices $\omega$ and $L$ explicitly in coordinates, separating the diagonal part from the rest (no summation implied):
\begin{align}
  \omega_{ij} &= \omega_i \delta_{ij}+\frac{K_{ij}}{b\pqty{a_i - a_j}^2},\\
  i L_{ij} &= i \dot a_i \delta_{ij}+ \frac{K_{ij}}{b\pqty{a_i - a_j} }.
\end{align}
The diagonal part of the \ac{eom} can now be written as
\begin{equation}
\ddot{a}_i - \frac{2}{b^2}\sum_{j\neq i} \frac{|K_{ij}|^2}{(a_i-a_j)^3} + \frac{R}{8}a_i+
V'(A)_{ii}
=0
.
\end{equation}
We already know that when $g_1 \neq 0$, $a_2 = -a_1$ and $a_3=0$. This means that we do not have three independent equations, but one equation and a set of consistency constraints for the matrix $K$:
\begin{align}
\label{LSM:SU3diagonalEOM}
  &\ddot a_1 - \frac{2}{b^2a_1^3} \pqty{ \frac{\abs{K_{12}}^2}{8} + \abs{K_{13}}^2} + \frac{R}{8} a_1 + 4g_0 a_1^5 =0, \\
  &
  \abs{K_{13}}^2 = \abs{K_{23}}^2.
\end{align}
We can choose a gauge in which $K$ is real (observe that only the absolute value of $K$ enters the Hamiltonian~\eqref{eq:linear-sigma-Hamiltonian}) and we find that $K$ only depends on two parameters and can be written as
\begin{equation}
  K =  \rho
  \begin{pmatrix}
    0                               & \cos \theta                   & \frac{1}{2\sqrt2} \sin \theta \\
    \cos \theta                     & 0                             & \frac{1}{2\sqrt2} \sin \theta \\
    \frac{1}{2\sqrt2} \sin \theta   & \frac{1}{2\sqrt2} \sin \theta & 0 \end{pmatrix} =
  \rho \pqty{ \cos\theta \lambda_1+\frac{1}{2 \sqrt 2}\sin\theta \pqty{\lambda_4 + \lambda_6}}.
\end{equation}
How are $\rho$ and $\theta$ related to the conserved charge density $J_0$? First, we observe that because of \(SU(N)\) invariance only the eigenvalues of $J_0$ are physical. Since the matrices $K$ and $J_0$ are similar, they have the same spectrum:
\begin{equation}
\label{LSM:J0_Eigenvalues}
  \spec(J_0) =  \spec(K) = \left\{ \rho \cos^2\frac{\theta}{2},\ \rho\left( -\cos^2\frac{\theta}{2}+\sin^2\frac{\theta}{2}\right), -\rho\sin^2\frac{\theta}{2} \right\}.
\end{equation}
The conservation of \(J_0\) then implies that \(\dot \theta = \dot \rho = 0\) and then in turn \(\dot K = 0\). The charge density $J_0$ can be interpreted as a vector in the space spanned by the maximal torus of the symmetry group, which in the case of $SU(3)$ is $U(1) \times U(1)$ and can be parametrized by two parameters, its modulus and an angular variable. $\rho$ corresponds to the modulus and acts as a charge density, and the angle which parametrizes the embedding of $J_0$ into the maximal torus corresponds to $\theta$.

\medskip

We have found that, in the \(SU(3)\) case, a homogeneous solution to the \ac{eom} can only have one fixed charge density \(\rho\). %
This is an important result.
For two independent fixed charges, no homogeneous solution exists.
This is the same situation that was encountered in~\cite{Alvarez-Gaume:2016vff}.\footnote{This is not true in general for a matrix model. The existence of a single Abelian charge discussed here is particular to the case of $SU(3)$.
Models with symmetry $SU(N)$ with $N>3$ do not have this property and can give rise to more general fixed points with more than one fixed charge.}
From now on, we will use \(\rho \gg 1 \) as an expansion parameter.

\medskip
We can now come back to the diagonal \ac{eom} \eqref{LSM:SU3diagonalEOM}.
Given the form of the Hamiltonian in Eq.~(\ref{eq:linear-sigma-Hamiltonian}),
requiring the lowest energy homogeneous solution means $\Pi_A=0$, hence $\dot A=0$.
Under this additional assumption the determining equation for $a_1 = \frac{v}{\sqrt 2}$ becomes
\begin{equation}
  - \frac{\rho^2}{b^2  v^3} + \frac{R}{8} v + g_0 v^5 = 0 .
\end{equation}
Solving this perturbatively in terms of $\rho$ we find for the amplitude $v$ %
\begin{equation}
  \label{eq:radial-ground-state}
  v = \frac{1}{2} \pqty{\frac{-R + \sqrt{R^2+256 g_0(\rho/b)^2}}{g_0}}^{1/4}
  =
  \frac{(\rho/b)^{1/4}}{g_0^{1/8}}
  \pqty{ 1 + \order{\frac{1}{\rho}}}
  .
\end{equation}

\paragraph{Calogero--Moser.}

A special case in the $A_2$ theory arises for $\tan \theta = 2\sqrt 2$, when
\begin{equation}
  K =  \frac{\rho}{3} \begin{pmatrix}0 &1 & 1\\ 1& 0 & 1 \\ 1 & 1& 0 \end{pmatrix} =   \frac{\rho}{3} \pqty{ \lambda_1 +\lambda_4+\lambda_6 } .
\end{equation}
Then, the Hamiltonian for the homogeneous system is of Calogero--Moser type.
In fact, starting from any $A_{N-1}$ Lagrangian \eqref{eq:L}, the Calogero--Moser Hamiltonian is given by Eq.~\eqref{eq:linear-sigma-Hamiltonian} at the homogeneous solution,
\begin{equation}
  \mathcal{H} =  \frac{1}{2} \sum_i \pi_{i}^2 + \frac{1}{
        2(N b)^2}\sum_{i\neq j} \frac{\rho^2}{(a_i-a_j)^2} +
  V(a_1,...,a_N)
  ,
\end{equation}
where all $\abs{K_{ij}} = \rho/N$ $\forall i\neq j$ (see~\cite{Polychronakos:2006nz} for a comprehensive review).
It describes a set of $N$ particles with the same charge density $\rho/\pqty{Nb}$ repelling each other in an attractive potential $V(a_i)$.

This gives us an intuitive understanding for the effect of fixing the charge.
Fixing $\rho\neq 0$, we see that in the homogeneous ground state, the eigenvalues of $\Phi$ do not collapse to the origin as one would expect in the $\Phi^6$ potential, but are distributed on the line like particles in the ground state of a Calogero--Moser model.
This behavior is only precise in the special case $\tan \theta = 2\sqrt 2$, but the qualitative picture remains the same for any value of $\theta$.

It should not be surprising that we have recovered a translation-invariant but non-Lorentz-invariant model, since we work at fixed charge, \emph{i.e.} in sectors where the Lorentz invariance is broken.

\paragraph{The non-diagonal part of the EOM.}
Let us now consider the non-diagonal part of the \ac{eom},
\begin{equation}
  \dot L_{ij} = i \comm{L}{\omega}_{ij} ~,~~  i \neq j.
\end{equation}
This implies charge conservation:
commuting both sides of the \ac{eom} in Eq.~\eqref{eq:eom} with $A$ (making its diagonal part drop) and invoking the Jacobi identity, we find
\begin{equation}
\label{EulerArnold_LaxForm}
[\dot L , A ] = i \left[[L,\omega],A \right] \overset{\text{Jacobi}}{=} i \left[[L,A],\omega \right] + i \left[L, [\omega,A] \right]
\overset{\eqref{def:LaxMatrix}}{=}
i \left[[L,A],\omega \right]  - \left[ L , \dot A \right]
.
\end{equation}
Using that the Lax form of the $K$ matrix in the $A_2$ algebra is
\begin{equation}
  \label{K_L_relation}
  K = - i b \comm{L}{A},
\end{equation}
we can rewrite Eq.~\eqref{EulerArnold_LaxForm} as
\begin{equation}
  \label{eq:Arnold}
  \dot K = -i \comm{\omega}{K} ~ \Leftrightarrow \dot J_0 = 0.
\end{equation}
This is the {Euler--Arnold equation} for the generalized rigid body.
Charge conservation follows from the off-diagonal part of the \ac{eom} in Lax form and it is independent of the potential $V(A)$.\footnote{As in the vector model, this follows by varying the angular \ac{dof}.} %

We have seen that on the ground state, $\dot K=0$.
It follows that for generic values of $\theta$, $\omega$ commutes with \(K\) and, in the gauge that we have used until now, it must have the form
\begin{equation}
  \omega = \mu \begin{pmatrix}
    \frac{1}{2}\cos\theta        & \frac{1}{2}\cos\theta        & \frac{1}{\sqrt2} \sin \theta \\
    \frac{1}{2}\cos\theta        & \frac{1}{2}\cos\theta        & \frac{1}{\sqrt2} \sin \theta \\
    \frac{1}{\sqrt2} \sin \theta & \frac{1}{\sqrt2} \sin \theta & -\cos\theta \end{pmatrix}
  = \mu \bqty{\frac{\cos \theta}{2} \pqty{ \sqrt{3} \lambda_8 + \lambda_1 } + \frac{\sin \theta}{\sqrt{2}} \pqty{ \lambda_4 + \lambda_6 }  } ,
\end{equation}
in terms of 
\begin{equation}
\label{LSM:MuLambdaV_Relation}
\mu =\sqrt{\frac{R}{8} + g_0 v^4}
~.
\end{equation}
Again, the only gauge-invariant information is the spectrum:
\begin{equation}
  \spec(\omega) = \set{\mu , 0, - \mu } .
\end{equation}
This means that we can pick a different gauge where \(\omega \) is diagonal and is written as \(\omega = -i U^\dagger \dot U\) with
\begin{equation}
  U = e^{i \mu t h} U_0 ,
\end{equation}
where
\begin{equation}
  \label{eq:h-chemical-potential}
  h = \pmqty{ \dmat{ 1 , 0, -1}}
\end{equation}
and \(U_0\) is the matrix that diagonalizes both \(K\) and \(\omega\),
\begin{equation}
  \label{eq:diagonalizing-matrix}
  U_0 =
 \begin{pmatrix}
   \frac{1}{\sqrt 2} \cos\frac{\theta}{2} & \frac{1}{\sqrt 2} \cos\frac{\theta}{2} &  \sin\frac{\theta}{2} \\
-\frac{1}{\sqrt 2} &  \frac{1}{\sqrt2} &0 \\
-\frac{1}{\sqrt 2} \sin\frac{\theta}{2}  & -\frac{1}{\sqrt 2} \sin\frac{\theta}{2} & \cos\frac{\theta}{2}
 \end{pmatrix}.
\end{equation}

\paragraph{The ground state.}

We can now write the form of the general homogeneous ground state, collecting the results that we have found above:
\begin{equation}
\label{LSM:ClassicalSolutionFinal}
{\Phi(t)} = \Ad[U(t)] {A}  =
  \Ad[e^{i \mu t h} U_0] {A}=\Ad[e^{i\mu th}
  ]\Phi_0,
\end{equation}
where \(h\) and \(U_0\) are defined respectively in Eq.~(\ref{eq:h-chemical-potential}) and Eq.~(\ref{eq:diagonalizing-matrix}), and
 \(\ev{A} = \frac{v}{\sqrt 2}\lambda_3\). The constant part of the \ac{vev} is now given by
\begin{equation}
\label{LSM:Phi0}
\Phi_0= \Ad[U_0]{A} =  \frac{v}{\sqrt 2}\Ad[e^{i \frac{\theta}{2}\lambda_5}] \lambda_1 = \frac{v}{\sqrt 2}
\begin{pmatrix}
0 & -\cos\frac{\theta}{2} & 0\\
-\cos\frac{\theta}{2} & 0 & \sin\frac{\theta}{2} \\
0 & \sin\frac{\theta}{2} &  0
\end{pmatrix},
\end{equation}
with \(v\) being the constant solution to the radial equation~(\ref{eq:radial-ground-state}).
All in all, in the chosen gauge we thus have
\begin{equation}
  \label{eq:A2-homogeneous-solution}
 { \Phi(t)} = \frac{v}{\sqrt{2}}
    \begin{pmatrix}
    0                           & -e^{i\mu t} \cos\frac{\theta}{2} & 0                         \\
    -e^{-i\mu t} \cos\frac{\theta}{2} & 0                          & e^{i\mu t} \sin\frac{\theta}{2} \\
    0                           & e^{-i\mu t} \sin\frac{\theta}{2} & 0                         \\
  \end{pmatrix} .
\end{equation}
The only thing that remains to do is to relate \(\mu\) to the charge density \(\rho\).
This is done by using the definition of the conserved current:
\begin{equation}
  J_0 = i B\comm{\Phi}{\dot \Phi}
  = b \mu v^2 \pmqty{
    \dmat{
      \cos^2\frac{\theta}{2}, -\cos^2\frac{\theta}{2}+\sin^2\frac{\theta}{2}, -\sin^2\frac{\theta}{2}
    }
  },
\end{equation}
whence (by comparing to \eqref{LSM:J0_Eigenvalues}) we conclude immediately that
\begin{equation}
\label{LSM:RelationMu_Rho}
  \mu = \frac{\rho}{b v^2} =
  \frac{\rho}{4b } \pqty{\frac{-R + \sqrt{R^2+256 g_0(\rho/b)^2}}{g_0}}^{-1/2}
  =
  g_0^{1/4} \pqty{\frac{\rho}{b}}^{1/2}
  \pqty{ 1 + \order{\frac{1}{\rho}}}.
\end{equation}
It is important to realize that there is only one control parameter, namely the conserved charge density \(\rho\).
It will however be convenient in the following to use either \(v = \order{\rho^{1/4}}\) or \(\mu = \order{\rho^{1/2}}\) to write asymptotic expansions in the limit of \(\rho \gg 1\).

Plugging the solution \eqref{eq:A2-homogeneous-solution} into the Hamiltonian \eqref{eq:linear-sigma-Hamiltonian},
we can calculate the condensate energy density
\begin{equation}
E_0 = \frac{(\mu v)^2}{2} + \frac{R v^2}{16} + \frac{g_0 v^6}{6}.
\end{equation}
Using the relations~\eqref{eq:radial-ground-state} and \eqref{LSM:RelationMu_Rho}, it can be expressed entirely in terms of the charge density $\rho$:
\begin{equation}
  E_0 %
  =
  \frac23 \frac{\lambda^{1/4}}{b^{3/2}} \rho^{3/2}
  +
  \frac{1}{16} \frac{R}{g_0^{1/4} \sqrt{b}} \sqrt{\rho}
   + \order{\frac{1}{\sqrt{\rho}}} .
\end{equation}
We can now use this result to calculate the leading contribution to the anomalous dimension. By the state-operator correspondence, the condensate energy on the sphere is the leading contribution to the anomalous dimension. Using that the Ricci scalar $R_{S^2}=2$ for a two-sphere of radius 1 and $Q = \rho V$, where $V=4\pi$ is the volume of the sphere, we eventually have:
\begin{equation}
\label{eq:anomDim}
 D(Q) =   \frac23 \frac{g_0^{1/4}}{b^{3/2}} \left(\frac{Q}{4\pi}\right)^{3/2}
  +
  \frac{1}{8} \frac{1}{g_0^{1/4} \sqrt{b}} \left(\frac{Q}{4\pi}\right)^{1/2}
   + \order{\frac{1}{\sqrt{Q}}} .
 \end{equation}
We see that we find the same universal behavior found in~\cite{Hellerman:2015nra,Alvarez-Gaume:2016vff}. In the next section, we will study the fluctuations  to find the corrections to this leading behavior.

\subsection{Fluctuations}
\label{sec:fluctuations}

\paragraph{Explicit symmetry breaking.}

Now that we have found an explicit expression for the fixed-charge homogeneous solution to the \ac{eom}, we want to quantize the fluctuations on top of it. It is convenient to write the field $\Phi$ as
\begin{equation}
  \Phi = \Ad[e^{i\mu th}]\widetilde\Phi,
\end{equation}
where $\widetilde\Phi$ contains both the constant \ac{vev} \(\Phi_0\) and the fluctuations.
Substituting this expression into the Lagrangian, we find that $\mu$ takes the role of a chemical potential for the $U(1)$ symmetry generated by $h$:
\begin{equation}
\label{LSM:LagrangianWithChemicalPotential}
  \mathcal{L} = \frac{1}{2} \Tr\left( \del_\mu\widetilde\Phi\del^\mu \widetilde\Phi\right) + i\mu \Tr\left( [\widetilde\Phi, \dot{\widetilde\Phi}]h\right) - \frac{\mu^2}{2}\Tr[h,\widetilde\Phi]^2 - V(\widetilde\Phi),
\end{equation}
thus explicitly breaking the $SU(N)$ symmetry to a subgroup $H$ that contains the centralizer of $h$:
\begin{equation}
\label{LSM:CentralizerOfh}
H \supseteq  \mathcal{C}_G(h) = \set{ g \in G | \Ad[g]h = h }.
\end{equation}
This is consistent with the general observation in~\cite{Alvarez-Gaume:2016vff} that the quantum Hamiltonian corresponding to a fixed-charge classical system has a chemical potential term.

In our case, $N=3$ and $h=\text{diag}(1,0,-1)$, so we are left with the maximal torus $U(1)\times U(1)$ of $SU(3)$. We will see in the following that this is actually too restrictive and that at leading order in $\rho$, there is an accidental symmetry enhancement to $U(2)$.

\paragraph{Accidental symmetry enhancement and spontaneous breaking.}

On the fixed-charge ground state $\ev{\Phi(t)}$, the field $\widetilde\Phi$ develops a constant \ac{vev} $\ev{ \widetilde \Phi} = \Phi_0$ which in general breaks the unbroken $H$ spontaneously to some subgroup $H'$. Goldstone's theorem tells us that the low-energy physics is described by $\dim(H/H')$ massless degrees of freedom. Even though the full theory is Lorentz invariant, we are considering a fixed-charge sector. This means that we break $SO(1,2)$ to $SO(2)$.
It follows that in general, we expect both relativistic and non-relativistic massless particles.
In particular for $N=3$, the $\Phi_0$ in Eq.~\eqref{LSM:Phi0} spontaneously breaks $U(1)^2$ to nothing, thus we are naively expecting two Goldstone \ac{dof}.

To explicitly investigate the fluctuations around the classical ground state found in Eq.~\eqref{LSM:ClassicalSolutionFinal},
one has to start with a coset parametrization of the form
\begin{equation}
  \Phi = \Ad[e^{i \mu t h}]
  \Ad[e^{i\frac{\theta}{2}\lambda_5}]
  \Ad[ e^{i \hat \phi_a T^a}]
  \pqty{ \frac{v}{\sqrt2}\lambda_1 + \hat \Phi_r},
\end{equation}
where \(\hat \Phi_r\) summarizes the ``radial'' directions that commute with %
$\lambda_1$,
and \(T^a\) are the remaining generators of the algebra, \emph{i.e.} \(\ev{T^a } = \set{g \in \mathfrak{g} | \comm{g}{\lambda_1} \neq 0}\).
In general, \(\hat \phi_a\) will be a Goldstone if the corresponding \(T^a\) commutes with \(h\).
On the other hand, a stable expansion around $\ev{\Phi(t)}$ always implies that the radial modes in $\hat \Phi_r$ are massive.

Let us separate, for the moment arbitrarily, the \(T^a\) into \((\Sigma^\alpha, N^b)\) where the \(\Sigma^\alpha\) generate the \(A_1\) subalgebra that contains \(h\), \emph{i.e.}
\begin{align}
  \Sigma^1 &= \lambda_4, & \Sigma^2 &= \lambda_5, & \Sigma^3 &= \frac{\sqrt{3}}{2} \lambda_8 + \frac{1}{2} \lambda_3 =  h~,
\end{align}
and the \(N^b\) are the remaining generators in $\set{T^a}$,
\begin{align}
  N^1 &= \lambda_6, & N^2 &= \lambda_2, & N^3 &=\lambda_7.
\end{align}
In order to obtain diagonal kinetic terms in the field expansion, it is convenient to reformulate the coset parametrization up to corrections of higher order in \(\mu\):
\begin{equation}
  \Phi = \Ad[e^{i \mu t h}]
  \Ad[U_\pi]
  \Ad[U_\varphi]
  \pqty{ \frac{v}{\sqrt2}\lambda_1 + \hat \Phi_r} +\order{\mu^{-1}}
\end{equation}
with
\begin{align}
  U_\pi &= \exp( i  \frac{\pi_3}{v} \Sigma^3)
  \exp( i  \frac{\pi_1}{v} \Sigma^1 ) \exp ( i \left(\frac{\theta}{2}+\frac{\pi_2}{v}\right) \Sigma^2 ),  \\
  U_\varphi &= \exp( i \frac{\varphi_1}{2v} N^1 + i\frac{\varphi_2}{v} N^2 + i \frac{\varphi_3}{v} N^3),
\end{align}
where the normalization for the fields \(\pi\) and \(\phi\) is chosen to result in a canonical kinetic term.
We also decompose the radial fluctuations explicitly as
\begin{equation}
  \hat \Phi_r = \frac{1}{\sqrt{2}} \pqty{ r_1 \lambda_1 + r_2 \lambda_8} .
\end{equation}
Substituting this parametrization of the fluctuations into the Lagrangian and expanding to leading order in the charge --- which coincides with second order in the fields ---  we find (up to boundary terms):
\begin{equation}
  \label{LSM:QuadraticLagrangian}
  \begin{aligned}
    \mathcal L^{(2)} ={}& \frac{\mu^2 v^2}{2} - \frac{1}{6} g_0 v^6 - \frac{1}{16} R v^2 \\
    & + \frac{1}{2} \left(\sum_{i=1}^3(\partial_\mu \pi_i)^2 + \sum_{i=a}^3(\partial_\mu \varphi_a)^2 + (\partial_\mu r_1)^2 + (\partial_\mu r_2)^2\right) \\
    & + 2\mu r_1 \dot \pi_3 - 2 \mu \pi_1 \dot \pi_2 + 2\mu \left(\sin\theta \varphi_2  +2 \cos\theta \varphi_3 \right) \dot{\varphi_1}
    - 2 \sqrt3 \mu \sin \theta  r_3 \dot{\varphi_1} \\
    & + \frac{1}{2} \mu^2 r_1^2 + \frac32 \mu^2 \varphi_1^2 - \frac{1}{2} \mu^2 \cos^2\theta \varphi_2^2 + \mu^2 \sin2\theta \varphi_2 \varphi_3 + \left(\frac{1}{2} + \cos2\theta \right)\mu^2\varphi_3^2 \\
    & - \sqrt{3} \mu^2 \left(\sin^2\theta  \varphi_2 + \sin2\theta \varphi_3\right) r_2 + \frac32 \mu^2 \sin^2 \theta  r_2^2 + \\
    & - \left(\frac{5}{2} g_0 v^4 + \frac{1}{16} R \right)  r_1^2 - \left(\frac{5}{6}g_0 v^4 - 2 g_2 v^4 + \frac{1}{16} R \right)  r_2^2 + \order{1/v}.
\end{aligned}
\end{equation}
As expected, at this order, more fields have become massless.
Together with the \emph{bona fide} Goldstone \(\pi_3\), corresponding to the symmetry \(\pi_3 \to \pi_3 + \epsilon\) of the fixed-chemical-potential action, there are two approximate (in the sense of large charge) Goldstone fields \(\pi_1\) and \(\pi_2\) which together parametrize the \(U(2)/U(1) = SU(2)\) coset.
Physically, they relate vacua with the same condensate energy but different charge assignment (different $\theta$ in \eqref{eq:diagonalizing-matrix}).
This means that --- at leading order in $Q$ --- the spontaneous symmetry breaking pattern is $U(2) \to U(1)$ and we expect three massless \ac{dof}.
In total, we thus have
\begin{equation}
  \label{LSM:BreakingPattern_GenericTheta}
  SU(3) \xrightarrow{\text{explicit}} U(2) \xrightarrow{\text{spontaneous}} U(1).
\end{equation}

In order to study the low-energy physics, it is convenient to pass to a non-linear sigma model approach, which we obtain by integrating out all the massive \ac{dof} and describing the low-energy physics in terms of a field $U \in SU(2)$.
In this framework, it will also be easier to show the suppression of higher-derivative terms and quantum effects by $1/Q$.

Before doing this in Sec.~\ref{sec:non-linear-sigma-model}, we first derive the dispersion relations for the Goldstones in the linear sigma model framework and comment briefly on the massive modes.

\paragraph{Dispersion relations.}

Starting from the quadratic Lagrangian in Eq.~\eqref{LSM:QuadraticLagrangian} it is straightforward to read off the inverse propagator in momentum space $D^{-1}(k) $, which takes a block-diagonal form\footnote{The fields are ordered as
 $\set{r_1, \pi_3 , \pi_1 , \pi_2 , \varphi_1,  \varphi_2 ,  \varphi_3 , r_2}$.%
}
\begin{equation}
D^{-1}(k) = \begin{pmatrix}
D_{\pi}^{-1}(k)                  & 0                                                                                            \\
0                                & D_\varphi^{-1}(k)
\end{pmatrix}
~,
\end{equation}
with
\begin{equation}
  \label{LSM:GoldstonePropagator}
  \eval{D^{-1}_\pi(k)}_{r_1, \pi_3 , \pi_1 , \pi_2 } =
\begin{pmatrix}
 k^2-k_0^2+4 \mu ^2 - \frac R2     & -2 i k_0 \mu              & 0                                    & 0                           \\
2 i k_0 \mu                        & k^2-k_0^2               & 0                                    & 0                           \\
0                                & 0                       & k^2-k_0^2                            & 2 i k_0 \mu                   \\
0                                & 0                       & -2 i k_0 \mu                           & k^2-k_0^2
\end{pmatrix}
\end{equation}
and
\begin{equation}
  \label{LSM:SpectatorPropagator}
    D^{-1}_\varphi(k)%
  =
  \begin{footnotesize}
    \begin{pmatrix}
      k^2-k_0^2-3 \mu ^2           & 2 i k_0 \mu  \sin \theta       & 4 i k_0 \mu  \cos \theta                    & -2 i \sqrt{3} k_0 \mu  \sin \theta \\
      -2 i k_0 \mu  \sin \theta         & k^2-k_0^2+\mu ^2 \cos 2 \theta & -\mu ^2 \sin 2 \theta                       & \sqrt{3} \mu ^2 \sin^2\theta       \\
      -4 i k_0 \mu  \cos \theta         & -\mu ^2 \sin 2 \theta          & k^2-k_0^2-\mu^2 \pqty{ 1 - 2 \cos 2 \theta} & \sqrt{3} \mu ^2 \sin 2 \theta      \\
      2 i \sqrt{3} k_0 \mu  \sin \theta & \sqrt{3} \mu ^2 \sin ^2\theta  & \sqrt{3} \mu ^2 \sin2 \theta                & k^2-k_0^2 - m_{r_2}^2
    \end{pmatrix}
  \end{footnotesize},
\end{equation}
where \(m_{r_2}^2 = \mu^2 \pqty{\frac{4 g_2 }{g_0} + \frac{1}{6}+\frac{3}{2}  \cos 2 \theta} + R \left(\frac{g_2}{2g_0}-\frac{1}{12}\right) \).
Looking at the mass terms of the radial modes $r_1$ and $r_2$ it becomes immediately clear that any $R$-dependent contributions to the fluctuations are sub-leading.

\paragraph{Goldstone modes.}
Starting from $D^{-1}_\pi(k)$, which does not depend on the angle $\theta$ which describes the embedding of the fixed charge in the maximal torus,
\begin{itemize}
\item the first $2\times2$ block describes, after diagonalizing, a massive mode ($r_1$ to leading order) coupled to the universal relativistic Goldstone $\chi$:
\begin{align}
  \omega_{\chi} &= \frac{\abs{\mathbf{k}}}{\sqrt 2} + \order{\mu^{-1}}, &
                   \omega_{r_1} = 2\sqrt{2}\mu + \order{\mu^0}.
\end{align}
\item The second $2\times2$ sub-block of \eqref{LSM:GoldstonePropagator} describes the non-relativistic Goldstone sector,
\begin{align}
  \omega_{\pi}^- &= \frac{\abs{\mathbf{k}}^2}{2\mu} + \order{\mu^{-2}} , &
                                                           \omega_{\pi}^+ &= 2\mu + \frac{\abs{\mathbf{k}}^2}{2\mu} + \order{\mu^{-2}} ,
\end{align}
resulting from an accidental symmetry enhancement which happens at leading order in the charge \(\rho\).
\end{itemize}
These are precisely the same low-energy \ac{dof} that appear in the description of the \(O(4)\) vector model~\cite{Alvarez-Gaume:2016vff}.

The Casimir energy of the Goldstones gives the first correction to the conformal dimension Eq.~\eqref{eq:anomDim}; it is however easier to discuss this in the framework of the non-linear sigma model, which we do in Section~\ref{sec:non-linear-sigma-model}, where we also prove that the interactions are controlled by negative powers of the charge.

\paragraph{Massive modes.}
By diagonalizing $D^{-1}_\varphi(k)$ given in Eq.~\eqref{LSM:SpectatorPropagator} we determine the dispersion relations of the spectator fields:
\begin{align}
  \label{LSM:SpectatorModes_Dispersions}
  m_N^{1,2} &=\mu , &
  m_N^{\pm} &=\mu 
  \frac{ \sqrt{\delta +21 \pm \sqrt{\pqty{\delta - 6} ^2-54 \pqty{\delta - 9} \cos 2 \theta +567}}}{\sqrt{6}}
~,
\end{align}
where we have used the parameter \(\delta\) introduced in Eq.~\eqref{eq:alternative-couplings}.

First observe that the potential is bounded from below if \(\delta > 0\), which assures that the inner square root in \(m_N^{\pm}\) is real.
Moreover, \(m_N^+\) is always real and parametrically heavy, \(m_N^+ = \order{\mu}\).

We must be more careful with \(m_N^-\), though.
If \(0 < \delta < 6\), the argument of the square root can become negative and we get a stable mode (and a sensible large-charge expansion) only for some values of the angle \(\theta\), namely only if
\begin{equation}
  \cos(2\theta)  \leq \frac{3 - \delta}{ \delta - 9} ~.
\end{equation}
We find that even if the potential is bounded from below (\(\delta > 0\)), there exists a region in the space of the parameters \((g_1, g_2)\) where homogeneous fixed-charge solutions are possible only for certain ways of embedding the charge vector $J_0$ in the maximal torus of the symmetry group, parametrized by the angle $\theta$ (see Fig.\ref{fig:ExclusionRegions}).
\definecolor{allowedcolor}{RGB}{166,189,219}
\definecolor{semiallowedcolor}{RGB}{43,140,190}
\begin{figure}
  \centering
    \begin{tikzpicture}

      \draw[fill=semiallowedcolor, draw=none] (0,0) -- (-4, 2) -- (-4,3) -- cycle;
      \draw[fill=allowedcolor, draw=none] (0,0) -- (-4,3) -- (-4,3.5) -- (4,3.5) -- (4,-2) -- cycle;

      \draw[-latex] (-4,0) -- (4.5,0) node[below] {\(g_1\)};
      \draw[-latex] (0,-1) -- (0,4) node[left] {\(g_2\)};

      \draw[thick] (0,0) -- (-4, 2);
      \draw[thick] (0,0) -- (4, -2);
      \draw[thick] (0,0) -- (-4, 3);

      \node[] at (1.5,2) {allowed region};
      \node[text width=3cm] at (-5.25,2.5) {only for some charges};
      \node[] at (-3,-1) {unbounded potential};

    \end{tikzpicture}
  \caption{The validity regions in the $(g_1,g_2)$ plane.
    In the upper region (\(\delta > 6\)) the large charge expansion is valid for any fixed choice of $J(\rho,\theta)$.
    In the leftmost wedge (\(0< \delta < 6\)) there is a perturbative meaningful expansion only for certain values of $\theta$.
    The bottom region  (\(g_0 < 0\)) is not allowed because the scalar potential is not bounded.}
  \label{fig:ExclusionRegions}
\end{figure}
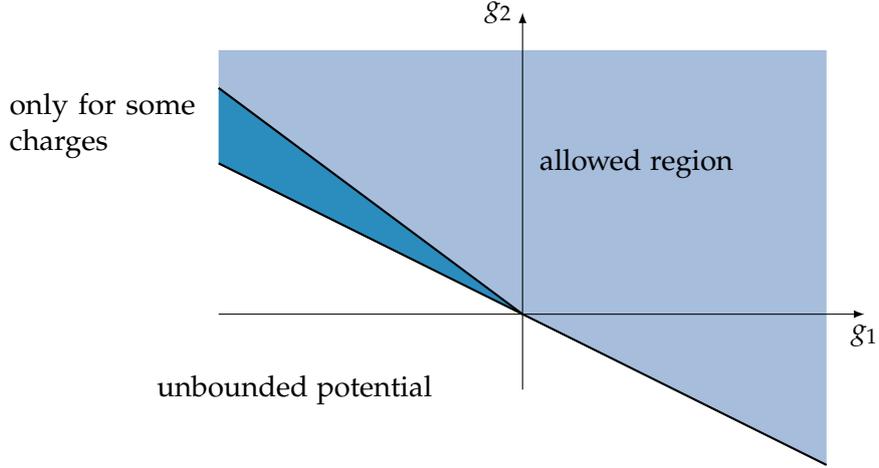

\paragraph{The $\theta=0$ special point.}
The accidental symmetry enhancement to $U(2)$ happens for \textit{generic} values of $\theta$.
In the special case of $\theta=0$, however, the off-diagonal \ac{eom}, Eq.~\eqref{eq:Arnold}, allow for yet another $\omega$, implying another possible choice for the chemical potential (respecting always charge quantization):%
\begin{equation}
h_{\theta=0} = \pmqty{ \dmat{ 1 , -1, 0}}.
\end{equation}
Performing the same analysis as above, we find that the symmetry-breaking pattern for this case is $U(1)^2 \to U(1)$.
No accidental symmetry enhancement happens: there is only one relativistic Goldstone $\chi$ and all other modes are parametrically massive. 
The non-linear sigma model for such a low-energy situation has already been discussed in~\cite{Hellerman:2015nra,Monin:2016jmo}.

Note that the same freedom exists in the vector model~\cite{Alvarez-Gaume:2016vff}, where the  homogeneous ground state of the $O(2N)$ model can be coupled via the chemical potential in different ways, resulting in the symmetry-breaking patterns $U(k) \to U(k-1)$ for any value of $k \leq N$.

\section{Non-linear sigma model}
\label{sec:non-linear-sigma-model}

The main purpose of the analysis of the previous section was to find the symmetry-breaking pattern resulting from studying the physics of the \(U(N)\)-symmetric model in a sector of fixed \(U(1)\) charge.

Now that we know that for \(N = 3\), at leading order in the charge density \(\rho\),  the pattern is \(U(2) \to U(1)\), we can integrate out all the massive \ac{dof} and write an effective action for the remaining Goldstones.
In fact, according to the general philosophy of low-energy effective actions, any Lagrangian that captures the right symmetries will describe the correct physics~\cite{Weinberg:1978kz}.

\paragraph{Effective action.}

We want to write an action for a field \(U\) in the coset \(U(2)/U(1) = SU(2)\) which is approximately scale-invariant, \emph{i.e.} that only contains terms of dimension three and respects a \(SU(2)_L \times SU(2)_R\) symmetry.
The action will contain derivatives of \(U\) and terms of the type
\begin{equation}
\label{NLSM:DelU_Def}
  \norm{ \del U} = \sqrt{\Tr( \del_\mu U^\dagger \del^\mu U )} ,
\end{equation}
which we can think of as resulting from integrating out the massive \ac{dof}.
More precisely, the action will have the form of an infinite sum of terms with arbitrary derivatives of \(U\) in the numerator and only powers of \(\norm{\del U}\) in the denominator.

In order to make this effective Wilsonian action useful, we will expand it around the fixed-charge ground state of Sec.~\ref{sec:homogeneous-ground-state}, so that \(\del_0 U = \order{\mu}\).
The analysis of the leading-order terms is then analogous to the one for the \(O(2)\) model discussed in~\cite{Hellerman:2015nra} and results in
\begin{equation}
  \label{eq:NLSM-LagrangianDensity}
  \mathcal{L} = \frac{c_1}{3 \sqrt 2} \norm{\del U}^3 - \frac{c_2}{\sqrt 2} R \norm{\del U} + \order{\mu^{-1}},
\end{equation}
where $R$ is the scalar curvature, and $c_1$ and $c_2$ are constants.

That the effective action for terms with positive \(\rho\)-scaling has only two parameters is consistent with the observation that the fluctuations around the ground state in the linear sigma model %
only depend on a linear combination of the couplings $g_1$ and $g_2$ and on the charge coupling $b$.
In the following we will identify the precise relationship between the two parametrizations by computing the energy of the ground state in the two descriptions (Eq.~\eqref{eq:two-parametrization-constants}).

\bigskip

The theory is invariant under the action of $SU(2)_L \times SU(2)_R$ and the corresponding Noether currents are
\begin{align}
\label{NLSM:NoetherCurrents}
  J^{L}_\mu &= 
  c_J(U) \pqty{i \del_\mu U U^\dagger}, 
  &
  J^{R}_\mu &= %
  c_J (U) \pqty{-i U^\dagger \del_\mu U },
\end{align}
where we have introduced $c_J(U)$ to abbreviate the frequently appearing factor 
\begin{equation}
  \label{NLSM:CurrentCoefficient}
  c_J(U) \equiv \frac{1}{\sqrt2}\pqty{ c_1 \norm{\del U} - c_2\frac{R}{\norm{\del U}}}
.
\end{equation}
It is also convenient to introduce the left-/right-invariant Maurer--Cartan forms\footnote{In the language of~\cite{arnold1999topological} $\omega_L$ is the angular velocity in the body and $\omega_R$ is the spatial angular velocity.}
\begin{align}
\label{AngularVelocities}
       \omega_L &\equiv -i U^\dagger \dot U  & \text{and} &&
       \omega_R &\equiv i \dot U U^\dagger 
        ~,
\end{align}
so that
\begin{equation}
  \norm{\del U}^2 = \Tr (\omega_L^2 - \abs{\nabla U}^2 ) = \Tr( \omega_R^2 - \abs{\nabla U}^2 )
~.
\end{equation}

Expressing the Lagrangian solely in terms of the angular velocity $\omega_L$, %
it is evident that $\omega_{L}$ and the zero-component of the Noether current $J^R_0$ in Eq.~\eqref{NLSM:NoetherCurrents} are in fact conjugate variables: 
\begin{equation}
J^R_0 = \left( \frac{\delta \mathcal{L}}{\omega_L}\right)^t = c_J(U) \omega_L,
\end{equation}
and the same applies for $\omega_{R}$ and $J^L_0$.
Thus, it is straightforward to write down\footnote{Equivalently, one could have defined $\pi = \left( \frac{\delta \mathcal{L}}{\delta \dot U^\dagger}\right)^t = \frac{1}{2 \sqrt 2} \left( c_1 \norm{\del U} + c_2\frac{R}{\norm{\del U}}\right)\dot U$ and Legendre transformed to $\mathcal H = \Tr(\pi \dot U^\dagger + \pi^\dagger \dot U) - \mathcal L$.} the Hamiltonian density, \emph{e.g.} in terms of $J^R_0$ current matrix:
\begin{equation}
  \label{NLSM:HamiltonianDensity}
  \begin{aligned}
    \mathcal{H} &=
    \Tr(J_0^R \omega_L) - \eval{\mathcal L}_{\omega_L= J_0^R / c_J(U)}
    \\
    &=
    \eval{
      c_J(U) \Tr \left(\omega_L^2\right) - \frac{c_1}{3 \sqrt2} \norm{\del U}^{3} +  \frac{c_2 R}{\sqrt 2} \norm{\del U} 
    }_{\omega_L= J_0^R / c_J(U)} .
  \end{aligned}
\end{equation}
\paragraph{Homogeneous ground state.}
Under the assumption of homogeneity in space, \emph{i.e.} $\nabla U=0$, we vary the action associated to~\eqref{eq:NLSM-LagrangianDensity} to derive the Euler--Lagrange \ac{eom}:
\begin{equation}
  \label{NLSM:EOMs}
  \dv{t} \omega_L = \dv{t} \omega_R =  0
~.
\end{equation}
We restrict our analysis to the $SU(2)$ case describing the symmetry-breaking pattern $U(2) \to U(1)$ that we have found in the previous section.
A convenient explicit parametrization is the one in terms of Euler angles:
\begin{equation}
  \label{NLSM:EulerParametrization}
  U_E(\pi_1, \pi_2, \pi_3) = e^{i\pi_3\sigma_3}e^{i\pi_1\sigma_2}e^{i\pi_2\sigma_3} =
  \begin{pmatrix}
    e^{i(\pi_3+\pi_2)} \cos\pi_1    & e^{i (\pi_3-\pi_2)} \sin\pi_1 \\
    -e^{-i (\pi_3-\pi_2)} \sin\pi_1 & e^{-i(\pi_3+\pi_2)} \cos\pi_1
  \end{pmatrix}
  ,
\end{equation}
where $\sigma_i$ are the Pauli matrices and the angles take the values $\pi_3 \in [0,\pi]$, $\pi_1 \in [0, \pi/2]$ and $\pi_2 = [0, 2\pi)$.
The matrices \(\omega_L\) and \(\omega_R\) are not independent since \(\omega_R = - U^\dagger \omega_L U\), and for \(SU(2)\) they share the same spectrum:
\begin{align}
  \spec(\omega_L) &= \spec(\omega_R) = \set{\pm \norm{\del U}} ,\\
  \norm{\del U}^2 &= \dot \pi_3^2 + \dot \pi_2^2 + \dot \pi_1^2 + 2 \cos(2 \pi_1) \dot \pi_3 \dot \pi_2 .
\end{align}
It follows that the \ac{eom} \(\dot \omega_L = \dot \omega_R = 0\) implies that
\begin{equation}
  \dv{t} \norm{\del U} = 0 .
\end{equation}
The energy is an increasing function of \(\norm{\del U}\) so, in order to minimize it keeping \(\omega_L \neq 0\) and \(\omega_R \neq 0\), we must have
\begin{equation}
  \begin{cases}
    \dot \pi_3 = \mu_1 = \text{const.} \\
    \dot \pi_2 = \mu_2 = \text{const.} \\
    \pi_1 = \text{const.}
  \end{cases}
\end{equation}
Then \(\omega_L\) takes the form
\begin{equation}
  \omega_L = \begin{pmatrix}
    -\mu_2 \cos2\pi_1 - \mu_1 & \mu_2  \sin2\pi_1 e^{2i \mu_1 t}\\
    \mu_2 \sin2\pi_1 e^{-2i \mu_1 t}   & \mu_2 \cos2\pi_1 +\mu_1
  \end{pmatrix}
  .
\end{equation}
This reduces Eq.~\eqref{NLSM:EOMs} to a single \ac{eom}:
\begin{equation}
  \mu_1 \mu_2 \sin 2 \pi_1 = 0 ,
\end{equation}
which is satisfied by
\begin{align}
\label{NLSM:CLassicalSolutions}
        \pi_1 &= 0 &\text{or} &&
        \pi_1 &= \frac{\pi}{2} &
        & \text {or} &&
        \mu_1 = 0 &\text{or} &&
        \mu_2 =0
        ~.
\end{align}
All these conditions eventually lead to the classical ground state
\begin{equation}
  \label{eq:SU2-ground state}
{U(t)}  = e^{-i \mu t \sigma_3} e^{i \pi_1 \sigma_2}.
\end{equation}
The unique coefficient \(\mu = \mu_1 \pm \mu_2\) characterizes the time-dependence of the classical solution, %
while \(\pi_1\) is a constant which is fixed by a gauge choice.

Once more we find that if we restrict ourselves to homogeneous solutions we can only fix one \(U(1)\) action (here the left and right actions are identified).
Obviously there are more general solutions where \(\mu_1\) and \(\mu_2\) are independent, but they will not be homogeneous.
Solutions of the type \(e^{i \mu_1 t \sigma_3} e^{i \pi_1(x) \sigma_2} e^{i \mu_2 t \sigma_3}\) have been recently discussed in~\cite{Hellerman:2017efx}.

If we pick \(\ev{\pi_1} = \pi /2 \), the solution representing our \ac{vev} takes the form
\begin{equation}
{U(t)} =
  \begin{pmatrix}
  \label{NLSM:ClassicalGroundState}
    0 & e^{-i \mu t} \\
    -e^{i \mu t} & 0
  \end{pmatrix}
\end{equation}
and the Noether currents on this classical ground state are diagonal:
\begin{equation}
\label{NLSM:NoetherCurrentsVEV}
{J_0^L} = {J_0^R} = \mu^2 \pqty{ c_1 - \frac{c_2 R}{2 \mu^2}} \sigma_3,
\end{equation}
where \(\sqrt{2} \mu = \norm{\del {U(t)}}\).
It is natural to fix the charge density for the adjoint action
\begin{equation}
\label{NLSM:AdjointCurrent}
  {J_0} = {J_0^L + J_0^R} = \rho \sigma_3
\end{equation}
and use \(\rho \gg 1\) as the controlling parameter or, equivalently, expand in powers of  \(\mu^2 = \pqty{ \rho + c_2 R}/\pqty{2c_1} = \order{\rho}\).

\paragraph{Fluctuations.}
We can now study the quantum problem, \emph{i.e.} the dynamics of the fluctuations over the solution in Eq.~(\ref{eq:SU2-ground state}).
It is convenient to parametrize the generic element $U$ starting from the gauge \(\ev{\pi_1} = \pi/4\) and write:
\begin{equation}
\label{NLSM:EulerParametrizaion}
  \begin{aligned}
  U &= U_E\left(\frac{\pi}{4} + \frac{\hat \pi_1}{\sqrt{2 c_1 \mu}}, \frac{\hat\pi_2}{\sqrt{2 c_1 \mu}},-\mu t + \frac{\hat \pi_3}{\sqrt{4 c_1 \mu}}\right) \\
  &= \exp[i\pqty{-\mu t + \frac{\hat \pi_3}{\sqrt{4 c_1 \mu}}} \sigma_3] \exp[i \pqty{\frac{\pi}{4} + \frac{\hat \pi_1}{\sqrt{2 c_1 \mu}}} \sigma_2] \exp[i \pqty{\frac{\hat\pi_2}{\sqrt{2 c_1 \mu}}}\sigma_3],
\end{aligned}
\end{equation}
where the normalization of the fluctuating fields is chosen such that when expanding the effective action~\eqref{eq:NLSM-LagrangianDensity}, the kinetic terms are canonical.
Expanding\footnote{We omit the hat for ease of notation.} at leading order in \(\mu\) we find:
\begin{equation}
  \label{eq:nlsm-expanded-lagrangian}
  \begin{aligned}
  \mathcal{L} ={}& \frac{2}{3}c_1 \mu^3 - \mu c_2 R \\
  &+ \frac{1}{2} \dot \pi_3^2- \frac{1}{4} (\nabla\pi_3)^2  \\
  &+ \frac{1}{2} \dot \pi_1^2 - \frac{1}{2} (\nabla\pi_1)^2 + \frac{1}{2} \dot \pi_2^2- \frac{1}{2} (\nabla\pi_2)^2 + 2 \mu \pi_1 \dot \pi_2 \\
  &+ \frac{2}{3c_1} \pi_1^3 \dot \pi_2 + \order{\mu^{-1/2}}.
\end{aligned}
\end{equation}
Note that in this case the expansion in \(\mu\) does not coincide with the expansion at quadratic order in the fields, because of the quartic interaction \(\pi_1^3 \dot \pi_2\). We will see that once the fields are rewritten in terms of the canonical oscillators that diagonalize the Hamiltonian, this term ends up being negligible.

Let us consider the various constituents of the action separately. We have
\begin{itemize}
\item a constant term with two contributions of order $\order{\rho^{3/2}}$ and $\order{\rho^{1/2}}$.
  This is %
  related to the energy of the ground state, which gives the dominant contribution in the large-$\rho$ expansion.\item a relativistic massless field $\pi_3$ with dispersion relation $\omega = \frac{1}{\sqrt2}k + \order{\rho^{-1/2}}$. This is the first contribution of order $\order{\rho^0}$ that we encounter and it is precisely the same dominating term that appears in the $O(N)$ vector model. Its contribution to the energy is due to the Casimir effect and for the unit two-sphere $\Sigma=S^2$, it is $c_0 = - 0.093$. This is the only quantum correction which is not controlled by the large charge.
\item a pair of fields $\pi_1$ and $\pi_2$ which are coupled via a quadratic term $\pi_1\dot\pi_2$ and a quartic term $\pi_1^3\dot\pi_2$.
\end{itemize}
Let us now concentrate on the latter terms.
If we limit ourselves to quadratic order in the fields, we can write the inverse propagator
\begin{equation}
  \eval{\D^{-1}(k)}_{\pi_1,\pi_2} = \begin{pmatrix}
    \mathbf{k}^2 - k_0^2 & -2ik_0 \mu \\
    2 i k_0 \mu            & \mathbf{k}^2
  \end{pmatrix},
\end{equation}
which we recognize as describing a massless complex scalar field $\varphi = \frac{1}{\sqrt2}\left(\pi_1 + i\pi_2\right)$ in presence of a chemical potential:
\begin{equation}
  \mathcal{L} = \pqty{\del_t + i\mu}\varphi^* \pqty{\del_t-i\mu} \varphi - \abs{\nabla \varphi}^2 - \mu^2 \abs{\varphi}^2.
\end{equation}
The corresponding quantum Hamiltonian
\begin{equation}
  \mathcal{H} = \varpi^* \varpi + \nabla \varphi^* \nabla \varphi + \mu^2 \varphi^* \varphi - \mu ( \varpi \varphi - \varpi^* \varphi^* )
\end{equation}
has already been diagonalized in~\cite{Alvarez-Gaume:2016vff} by going to momentum space and decomposing the canonical variables $\varphi, \varpi$ in terms of Heisenberg oscillators $a$ and $b$:
\begin{equation}
  \begin{aligned}
    \varphi(k) &= \frac{1}{\sqrt{2}\left(p^2+\mu^2\right)^{1/4}} \pqty{a(k)+b^\dagger(-k)},
    \\
    \varpi (k) &=  -i \frac{\left(p^2+\mu^2\right)^{1/4}}{\sqrt2} \pqty{a(k)-b^\dagger(-k)}
    .
  \end{aligned}
\end{equation}
From the expression for $\varphi(k)$  we read off the scaling of the real Goldstone fields, once expanded in the basis of canonical oscillators,
\begin{align}
  \pi_1(k) &= \frac{1}{\sqrt2} \left(\varphi(k) + \varphi^* (-k)\right) \sim \frac{1}{2\sqrt{\mu}} \left(a(k)+a^\dagger(-k)+b(k)+b^\dagger(-k)\right) , \\
  \pi_2(k) &= \frac{-i}{\sqrt2} \left(\varphi(k) - \varphi^* (-k)\right) \sim \frac{-i}{2\sqrt{\mu}} \left(a(k)-a^\dagger(-k)-b(k)+b^\dagger(-k)\right) ,
\end{align}
and the final form of the diagonalized quadratic Hamiltonian is
\begin{equation}
  \mathcal{H} = \pqty{\sqrt{k^2 + \mu^2} - \mu} a^\dagger(k) a(k) + \pqty{\sqrt{k^2 + \mu^2} + \mu} b^\dagger(k) b(k),
\end{equation}
which shows that in the large-charge limit, \(a\) is massless and \(b\) is massive.

\paragraph{Higher operators and quantum corrections.}

After having diagonalized the quadratic Hamiltonian, we are ready to move on to the interaction terms.

The first term appearing is the quartic interaction in the Lagrangian in Eq.~\eqref{eq:nlsm-expanded-lagrangian}: \(\pi_1^3 \dot \pi_2\).
Both the fields \(\pi_1\) and \(\pi_2\) are of order \(\order{\mu^{-1/2}}\) when expanded in terms of canonical oscillators. This means that \(\pi_1^3 \dot \pi_2\) gives a contribution of order \(\order{\mu^{-2}} = \order{\rho^{-1}}\) which is negligible with respect to the leading terms in the Hamiltonian. This justifies the choice of considering only up to quadratic terms in the expansion in the fields.

A similar reasoning can be applied to all the quantum and higher-derivative corrections to the effective action in Eq.~\eqref{eq:NLSM-LagrangianDensity}.
The intuitive way of understanding this is that since we are working in a sector of fixed charge \(Q\), we have an effective scale \(\mu\) which controls both the higher-derivative terms and the effective dimensionful couplings, thus bypassing one of the main technical hurdles of the standard formulation of the Wilsonian action for a second-order phase transition.

The final result is the same as in~\cite{Alvarez-Gaume:2016vff}.
The leading correction to the energy of the ground state comes from the Casimir energy of the Goldstones, which is the only term of order \(\order{\rho^0}\) and receives no further corrections.
More precisely, the only contribution comes from the relativistic field \(\pi_3\) and is the same as for the \(O(N)\) vector models.

Concretely, there are two leading contributions to the energy of the lowest state: the energy of the ground state and the Casimir energy \(E_C(\Sigma)\) for a massless boson with speed of light \(1/\sqrt{2}\) compactified on \(\Sigma\):
\begin{equation}
  E = \ev{\mathcal H} + E_C(\Sigma) = 
  \frac{4 c_1}{3}  \mu^3 + E_C(\Sigma)
  =
  \frac{1}{3}\sqrt{\frac{2}{c_1} } \rho^{3/2}  + \frac{c_2}{\sqrt{2c_1} } R \rho^{1/2} + E_C(\Sigma) + \order{\rho^{-1/2}}
  ,
\end{equation}
where in the last equality we have used \(\mu^2 = \pqty{\rho + c_2 R}/\pqty{2 c_1}\),  which follows from fixing the adjoint Noether current in Eq.~\eqref{NLSM:AdjointCurrent}.

\bigskip

Using the state-operator correspondence and choosing \(\Sigma = S^2\) we recover the formula for the conformal dimension of the lowest primary of charge \(Q\):
\begin{equation}\label{eq:confDImFull}
  D(Q) = \frac{c_{3/2}}{2 \sqrt{\pi}} Q^{3/2} + {2 \sqrt{\pi}}{c_{1/2}} Q^{1/2} - 0.093 + \order{Q^{-1/2}},
\end{equation}
where we used \(E_C(S^2) = -0.093\)~\cite{Monin:2016bwf}.
This expression is completely analogous to the one for the \(O(N)\) model.
The only difference is in the precise value of the coefficients \(c_{3/2}\) and \(c_{1/2}\) that cannot be computed in this framework but require a different non-perturbative analysis.\footnote{Comparing with the condensate energy Eq.~\eqref{eq:anomDim}, we can express the coefficients in Eq.~\eqref{eq:confDImFull} either in terms of \(g_0, b\) appearing in the linear sigma model of Sec.~\ref{sec:linear-sigma-model}, or in terms of \(c_1, c_2\) that were used in this section:
  \begin{align}
    \label{eq:two-parametrization-constants}
  c_{3/2} &= \frac{g_0^{1/4}}{6 \pi b} = \frac{1}{6 \pi \sqrt{2c_1}} ,
            &
              c_{1/2} &= \frac{1}{32 \pi \sqrt{b} g_0^{1/4}} = \frac{c_2}{2 \pi \sqrt{2 c_1}}.
\end{align}
}

\section{CCWZ formalism and fusion coefficients}
\label{sec:CCWZ}

The main result of this section is the calculation of a three-point function for our \ac{cft} in the limit of large charge. As in the previous section, we take advantage of the state-operator correspondence and map $\setR^3$ to $\setR_t\times S^2$ with the dilatation operator in $\setR^3$ identified with the time-translation operator (\emph{i.e.} the Hamiltonian) in $\setR_t\times S^2$.

\subsection{Spontaneously broken internal and space-time symmetries}

We want to reproduce the symmetry breaking pattern $SU(2) \to \Phi$ together with the breaking of the conformal group $SO(d+1,1)$:
\begin{equation}
  \label{CCWZ:SymmetryBreakingPattern}
  SO(d+1,1)\times SU(2) \to SO(d)\times D',
\end{equation}
where $D'$ is the combination of dilatations and internal rotations that remain unbroken in the fixed-charge sector. We introduce a non-coordinate basis $\hat e_a = e\indices{_a^\mu}\del_\mu$ and its inverse $\hat e^a =  e\indices{^a_\mu}\dd{x^\mu}$.
In terms of infinitesimal generators, we have
\begin{equation}
  \label{CCWZ:FullBreakingPattern}
  \begin{aligned}
    \text{broken generators}                    & :
    \begin{cases}
      B_i \equiv J_{0i}                         & \text{boosts}       \\
      D                                         & \text{dilations}    \\
      \sigma_\alpha                             & \text{internal global symmetries}
    \end{cases}                                                       \\
    \text{unbroken generators:}                 & :
    \begin{cases}
      P_a' = P_a + \mu \delta\indices{_a^0} \sigma_3 & \text{translations} \\
      ~J_{ij}                                   & \text{rotations},
    \end{cases}
  \end{aligned}
\end{equation}
where $D\equiv P_0$ is identified with the dilatation operator on the cylinder and the ordinary Pauli matrices $\sigma_\alpha$ act as the internal symmetry generators in our application.
 The $B_i$ denote the generators of broken Lorentz boosts, while $P_0'$, $P_i$ and $J_{ij}$ for $i,j=1,2$ parametrize the $D'\times \text{SO}(3)$  invariance of the vacuum state.

We are breaking scale invariance, which means that a dilaton will appear in the spectrum. Using the \ac{ccwz} prescription, we can introduce a representative of the full coset space,
\begin{equation}
  \label{CCWZ:FullCosetRepresentative}
  W =  e^{i y^a P_a } e^{i\sigma D}  e^{i \eta^i B_i} U_E(%
        \pi_1,\pi_2,\pi_3 ),
\end{equation}
with the internal $U_E$  given in \eqref{NLSM:EulerParametrization}.
The tangent space coordinates $y^a(x)$, $a=0,...,d-1$ (which transform under translations $P_a$) are generically taken as functions of the space-time coordinates $x^\mu$.
The dilaton $\sigma$, rapidities $\eta^i$ as well as the internal $\pi_1, \pi_2, \pi_3$ are the Goldstones associated to the breaking pattern~\eqref{CCWZ:FullBreakingPattern}.
They are however not independent \ac{dof} in the low-energy regime. We will eliminate this redundancy by imposing a set of inverse Higgs constraints.

The simplest way to write the effective action is to introduce a covariant derivative with respect to the space-time symmetries:
\begin{equation}
  \label{CCWZ:FullCovariantDerivative}
  \mathcal{D} = \dd{} +
  i \left( \hat e^a - \dd{y^a} + \Omega\indices{^a_b} y^b - A y^a \right) P_a
  + \frac{i}{2} \Omega^{ab} J_{ab} + i A D,
\end{equation}
where $\Omega^{ab}$ is the connection one-form,  $A$ is the gauge field for the dilatations, and $J_{ab}$ is gauge field for Lorentz transformations.
The connection one-form is gauged away by imposing that $T\indices{^a_{bc}}=0$ be torsionless. Then, at lowest order in the derivative expansion, $\Omega\indices{^a_b}$ is a function of the dreibein $\hat e\indices{^a}$ coupled to the dilaton gauge field $A_\mu$:
\begin{equation}
  \label{CCWZ:SpinConnection}
  \Omega\indices{^{ab}_\mu} = \frac{1}{2} \pqty{e^{a \nu} \pqty{\partial_\mu e\indices{^b_\nu} - \partial_\nu e\indices{^b_\mu}}
+ e\indices{^c_{\mu}} e\indices{^a_{\nu}}e^{b \lambda} \partial_\lambda e\indices{_c^{\nu}}
 - (a\leftrightarrow b) }
 - \pqty{e\indices{^a_\nu}e\indices{^b_\mu} - e\indices{^b_\nu} e\indices{^a_\mu}} A^\nu.
\end{equation}
Now we have the covariant derivative to define the Maurer--Cartan one-form for our coset representative~\eqref{CCWZ:FullCosetRepresentative}.
The idea is to introduce a set of derivatives for the Goldstones, which transform covariantly under \emph{all} the symmetries (including the spontaneously broken ones) in order to have a set of building blocks for invariant Lagrangians. Explicitly,
\begin{equation}
  \label{CCWZ:MaurerCartanForm}
  -i W^{-1} \mathcal{D} W =
  e^{-\sigma} \hat e^a \Lambda\indices{_a^b}
  \left( %
    P_b'
    + %
    \omega\indices{_b^\alpha}  \sigma_\alpha %
    + \nabla_b \sigma D + \nabla_b \eta^i B_i + \frac{1}{2} \Xi\indices{^{ij}_b} J_{ij} \right),
\end{equation}
where
\begin{itemize}
\item $\Lambda^d_{\phantom{d}b} \equiv \pqty{e^{i \eta^i B_i}}\indices{^d_b}$ is the Lorentz transformation given by the boosts, which is equivalently parametrized by the rapidities
  \begin{align}
    \label{RapidityDEF}
    \beta_i &=  \frac{\eta_i}{\eta} \tanh\eta, %
    &
      \eta &= \sqrt{\eta^i\eta_i}.
  \end{align}
  Explicitly:
  \begin{align}
    \label{eq:Lorentz-boosts-beta}
    \Lambda\indices{^0_0} &= \gamma = \cosh \eta , &
    \Lambda\indices{^0_1} &= -\gamma\beta_i ,&
    \Lambda\indices{^i_0} &= -\gamma\beta^i ,&
    \Lambda\indices{^i_j} = \delta\indices{^i_j} + \pqty{\gamma - 1} \frac{\beta^i \beta_j}{\beta^k\beta_k} .
  \end{align}
\item The covariant derivative for the dilaton $\sigma$ is
  \begin{equation}
    \label{CCWZ:DilatonCovariantDerivative}
    \nabla_b \sigma = e^\sigma e\indices{_d^\nu}\Lambda\indices{^d_b} \left(\partial_\nu \sigma + A_\nu \right).
  \end{equation}
\item The covariant derivative of the internal Goldstones is
  \begin{equation}
    \label{CCWZ:InternalDOFsCovariantDerivative}
    \omega_b
    =
    e^\sigma \Lambda\indices{^c_b} e\indices{_c^\nu} \omega_\nu = e^\sigma \Lambda\indices{^c_b} e\indices{_c^\nu} \left( -i U^\dagger \del_\nu U \right),
  \end{equation}
  or in components,
  \begin{equation}
    \label{CCWZ:SU2MatrixElements}
    \omega\indices{_b^\alpha} = \frac{1}{2}\Tr \left( \omega_b\sigma_\alpha \right),
  \end{equation}
  where the $\sigma_\alpha$ are generators of $A_1$ (\emph{i.e.} the Pauli matrices).
\item The covariant derivative $\nabla_b\eta^i$ and the connection $\Xi_b^{ij}$ include higher-derivative terms of the Goldstone fields $\eta^i$ and are negligible in the large-charge expansion.
\end{itemize}

\subsection{The inverse Higgs constraints}

According to the standard lore for the spontaneous breaking of internal symmetries, the number of independent Goldstone modes equals the number of broken generators.
On the other hand, when space-time symmetries are spontaneously broken, we can have in principle fewer physical Goldstone fields than broken generators (see \emph{e.g.}~\cite{Nicolis:2013lma}).

In Section~\ref{sec:fluctuations} we have derived the existence of three low-energy modes for the symmetry breaking pattern in Eq.~\eqref{CCWZ:SymmetryBreakingPattern} by analyzing the linear sigma model.
This means that of the fields we have used to initially define the coset in Eq.~\eqref{CCWZ:FullCosetRepresentative} and the covariant derivative Eq.~\eqref{CCWZ:FullCovariantDerivative} the dilaton $\sigma$, the boost Goldstones $\eta^i$, the gauge field for dilatations $A_\mu$ and the spin connection $\Omega_\mu^{ab}$ are redundant \ac{dof} and thus must be gauged away.
Since we are not interested in describing a theory of gravity, we should as a first step eliminate the corresponding dynamical \ac{dof}. Hence, we can impose
\begin{align}
  \label{CCWZ:GravityInverseHiggsConstraints}
  T\indices{^a_{bc}} &= 0
                       & \text{and} && \nabla_b \sigma &= 0 ~.
\end{align}
The torsionless condition eliminates the spin connection $\Omega^{ab}$ as independent \ac{dof} in favor of the vielbein $\hat e^a$, see Eq.~\eqref{CCWZ:SpinConnection}.
The latter condition in~\eqref{CCWZ:GravityInverseHiggsConstraints} eliminates (see Eq.~\eqref{CCWZ:DilatonCovariantDerivative}) the gauge field corresponding to dilatations:
\begin{equation}
\label{GaugeDilations}
\nabla_b \sigma = 0 \quad{\Rightarrow}\quad
A_\mu = - \partial_\mu \sigma
~.
\end{equation}

It is straightforward to supplement Eq. \eqref{CCWZ:GravityInverseHiggsConstraints} with  a set of left- and right-invariant (hence also invariant under the adjoint action) inverse Higgs constraints involving the internal covariant derivatives:
\begin{align}
\label{CCWZ:InverseHiggsConstraints}
 \Tr(\omega_b \omega^b ) &= \mu^2 %
 &\text{and} &&
 \Tr (\omega_i \omega_0 ) &= 0
 ~.
\end{align}
They can be summarized as 
\begin{align}
  \Tr (\tilde \omega_b \tilde \omega_0 ) &= 0
  & \text{with} &&
  \tilde \omega_b &= \omega_b - \frac{i}{\sqrt 2}\mu \delta\indices{^0_b} \Id.
\end{align}
The first constraint conveniently fixes the dilaton to
\begin{equation}
  \label{CCWZ:SigmaChi}
  \mu^2 e^{-2\sigma} = \Tr \left( \omega_\mu \omega^\mu\right) = \Tr \left(\partial_\mu U^\dagger \partial^\mu U\right)
\equiv \norm{\del U}^2~,
\end{equation}
in terms of the familiar $\norm{\del U}$ introduced in Eq.~\eqref{NLSM:DelU_Def}.

The other two conditions (which are compatible with the breaking of Lorentz invariance in the fixed-charge sector) are used to eliminate the Goldstones \(\eta^i\).
It is convenient to use the results of the previous section to parametrize \(\omega\).
Concretely, write \(U \in SU(2)\) as in the Euler parametrization of Eq.~\eqref{NLSM:EulerParametrizaion} where the expectation value and the fluctuations are separated.
In addition, we choose to work in the gauge specified in Eq.~\eqref{NLSM:ClassicalGroundState}.
After noting that for $R \times \Sigma$, 
\begin{align}
  \label{CCWZ:MatrixElementsScaling}
\omega_0 = e^{\sigma} \Lambda\indices{^d_0} e\indices{_d^\mu} \omega_\mu =
e^{\sigma} \gamma \mu \left(\sigma_3 + \order{\mu^{-1}}\right),
\end{align}
then, at leading order, the latter two inverse Higgs constraints imply
\begin{equation}
  \Tr( \omega_i \omega_0 ) = e^{\sigma}\gamma\mu \Tr(\omega_i \sigma_3) \pqty{1 + \order{\mu^{-1}}} = 0 ~\Rightarrow~ \omega\indices{_i^3 } = 0 .
\end{equation}
Using the explicit expression of \(\omega\) and \(\Lambda\) as function of \(\beta\) we find:
\begin{multline}
  \omega\indices{_i^3} = e^\sigma \Lambda\indices{^c_i} e\indices{_c^\nu} \omega\indices{_\nu^3}
  =
  e^\sigma \pqty{ \Lambda\indices{^0_i} e\indices{_0^\nu}  + \Lambda\indices{^j_i} e\indices{_j^\nu}  } \omega\indices{_\nu^3} \\
  = e^\sigma \pqty{ -\gamma \beta_i e\indices{_0^\nu}  + \bqty{ \delta\indices{^j_i} + \pqty{\gamma-1} \frac{\beta_i\beta^j}{\beta^2}} e\indices{_j^\nu}  } \omega\indices{_\nu^3} = 0 .
\end{multline}
The solution to leading order in $\mu$ of the two equations for $i=1,2$ is given by
\begin{equation}
  \label{CCWZ:BetaOmega_Relation}
  \beta_i =  \frac{e\indices{_i^\nu}\omega \indices{_\nu^3}}{e\indices{_0^\nu}\omega\indices{_\nu^3}} .
\end{equation}
which is well-defined since $\eval{\omega\indices{_\nu^3}}_{\nu = 0} \ \sim \mu \neq 0$.

\bigskip

After having imposed the inverse Higgs constraints, we have a set of independent low-energy \ac{dof}.
Moreover, we are not interested in deformations of the coset metric, apart from the dilaton which is fixed by the constraint in Eq.~\eqref{CCWZ:SigmaChi}; this fixes also the dreibein \(\hat e^a\).
The upshot is that the only remaining \ac{dof} are the Goldstones for the internal symmetry that parametrize \(\omega\).
This is of course consistent with our analysis of Sec.~\ref{sec:non-linear-sigma-model}.
In the next section we will see how the precise form of the Lagrangian in Eq.~\eqref{eq:NLSM-LagrangianDensity} is recovered in this formalism.

\subsection{The non-linear sigma-model re-derived}

According to the \ac{ccwz} prescription, the invariant action in $d$ space-time dimensions generically has the form
\begin{equation}
  S = \int \dd[d]{x} \mu^d \det(e^{-\sigma} \Lambda\indices{_a^{b}} \hat e^a ) F(\omega_a, R\indices{^{ab}_{cd}},\Xi\indices{_b^{ij}}) .
\end{equation}
Here we recognize the coset dreibein \(e^{-\sigma} e\indices{^a_\mu} \Lambda\indices{_a^{b}}\). $F$ is a dimensionless scalar function of the remaining building blocks reviewed in the preceding section, \emph{i.e.} the internal Goldstone covariant derivatives $\omega_a$,  the curvature field strengths $R$ and the connection $\Xi$.
Let us consider the two factors separately.

For the invariant measure we can write
\begin{equation}
  \dd[d]{x} \mu^d \det(e^{-\sigma}\Lambda_a^{~b} \hat e^a ) = \dd[d]{x} \det \hat e \dd[d]{x}  \mu^d e^{-d \sigma}
\end{equation}
and, imposing the inverse Higgs constraint in Eq.~\eqref{CCWZ:SigmaChi},
\begin{equation}
  \det \hat e \dd[d]{x}  \mu^d e^{-d \sigma} = \dd{t} \dd{\Sigma} \norm{\del U}^3.
\end{equation}
Thanks to our choice of inverse Higgs constraint in Eq.~\eqref{CCWZ:SigmaChi}, the coset geometry is completely expressed in terms of the geometry of the surface \(\Sigma\) and the norm \(\norm{\del U}\).
We will use this fact to simplify the analysis of the function \(F\).

Having imposed the inverse Higgs constraint, it is easy to see that \(F\) is only function of \(\omega\) and the curvature invariants of the surface \(\Sigma\). %
Moreover, at fixed charge, we have a scale \(\mu\) that suppresses the derivative terms.
This implies that, at leading order in \(\mu\), the function \(F\) must have the form
\begin{equation}
  F = \frac{c_1}{3\sqrt 2} - \frac{c_2}{\sqrt 2} \frac{R}{\norm{\del U}^2} + \order{ \mu^{-3}},
\end{equation}
where \(c_1\) and \(c_2\) are constants and \(R\) is the scalar curvature of \(\Sigma\).

All in all, we have reproduced the classical $\sigma$-model of Section~\ref{sec:non-linear-sigma-model}:
\begin{equation}
  S = \int \dd{t} \dd{\Sigma } \pqty{ c_1 \norm{\del U}^3 + c_2 \norm{\del U} R } + \order{\mu^{-1}} .
\end{equation}

\subsection{The three-point function}

So far, we have just introduced a reformulation of our previous result. The advantage of this formalism is that if we take $\Sigma=S^2$ and use the state-operator correspondence, we have a direct way of reconstructing operators of fixed charge and dimension (\emph{i.e.} transforming linearly under the broken group) in terms of the Goldstone \ac{dof}.

In our case, we follow the treatment in~\cite{Monin:2016jmo} and start from a representation of the unbroken $SO(2)$ generated by $J_{12}$ to define a field $\Phi$ that transforms linearly in a representation
\begin{equation}
  \kappa
  \left( \text e^{i\sigma D}\text e^{i \eta^i B_i} 
    \text e^{i \pi_3 \sigma_3}
  \right) 
\end{equation}
of the broken group. It tells us that a scalar operator of fixed dimension $\delta$ and internal charge $q$ is written (up to a multiplicative constant) as
\begin{equation}
  \mathcal O_{q,\delta} \propto  \mu^\delta e^{i\delta D} e^{i\pi_3\sigma_3} \pqty{1 + \order{\mu^{-1}}} ,
\end{equation}
where the factor $\mu^\delta$ is needed to give $\mathcal O_{q,\delta}$ the right dimension. Using the inverse Higgs constraint, we get
\begin{equation}
  \mathcal O_{q,\delta} = C \norm{\del U}^\delta  e^{i \pi_3  q} \pqty{1 + \order{\mu^{-1}}},
\end{equation}
where $C$ is a dimensionless constant. $\norm{\del U}$ contains all the Goldstone \ac{dof}. At leading order, the result is the same as the one found in~\cite{Monin:2016jmo}, which is not surprising since the authors describe a $U(1)$ symmetry breaking. Once more, the leading contribution in $\mu$ to the low-energy physics in our model comes precisely from the universal $U(1)$ relativistic Goldstone. 

We can now compute the three-point fusion coefficient for three primary operators $\mathcal{O}_{Q,\Delta_1}$, $\mathcal{O}_{-Q-\delta,\Delta_2}$ and $\mathcal{O}_{q,\delta}$ in the limit of $Q\gg 1$ to find that the leading contribution scales as $Q^{\delta/2}$:
\begin{equation}
  c_{Q+q,q,Q}= \frac{C_q}{c_1^{\delta/2}} Q^{\delta/2} \left( 1 + \order{Q^{-1/2}} \right),
\end{equation}
where $C_q$ is a function of the charge $q$ alone which we cannot compute. The effect of the non-relativistic Goldstones is sub-leading, but can be computed similarly.

\section{Conclusions}
\label{sec:conclusions}

Wilsonian actions are often of little practical use due to the infinitely many possible terms that appear in them, compatibly with the symmetries of the system. When however studying a model in a sector of large global charge $Q$, most of these terms are suppressed by inverse powers of $Q$, turning the Wilsonian effective action into a useful and useable object which admits a perturbative expansion in $1/Q$. In this paper, we have successfully applied the large-charge method to matrix models in $2+1$ dimensions, going beyond the vector models discussed so far in the literature.

\medskip

Owing to their relation to the \(\CP{N-1}\) model, which is under intensive investigation in the condensed matter community, $SU(N)$ matrix models make for an interesting object of study.
We have focused on the special case of $SU(3)$ whose algebraic structure is more tractable than the one of the cases with higher rank. We have concentrated on a homogeneous ground state which appears for one fixed charge and determined the associated symmetry-breaking pattern. As expected, we found that also in this case, the interaction terms are suppressed with $1/Q$.  Moreover, the formula for the anomalous dimension retains the same universal structure found in~\cite{Hellerman:2015nra,Alvarez-Gaume:2016vff}, the constant term being the same as in the vector model:
\begin{equation}
    D(Q) = \frac{c_{3/2}}{2 \sqrt{\pi}} Q^{3/2} + {2 \sqrt{\pi}}{c_{1/2}} Q^{1/2} - 0.093 + \order{Q^{-1/2}}.
\end{equation}
We also have calculated explicitly the fusion coefficients using the \ac{ccwz} formalism discussed in~\cite{Monin:2016jmo}, and found the same scaling as for the \(O(2)\) model:
\begin{equation}
  c_{Q+q,q,Q}= \frac{C_q}{c_1^{\delta/2}} Q^{\delta/2} \left( 1 + \order{Q^{-1/2}} \right).
\end{equation}
These two results are the same as those found in the literature for simpler cases due to an Abelianization which takes place at leading order in the charge.
The physics of the subleading non-relativistic Goldstone fields deserves further investigation.

\medskip
We also observe behaviors that do not occur in the class of vector models. On the one hand, we find that we cannot fix a generic $U(1)$ charge for all admissible values of the parameters in the effective potential.
On the other hand, we find that at leading order, there is a symmetry enhancement leading to a richer symmetry breaking pattern  than we would have naively expected.

\bigskip

For a special choice of the embedding angle $\tan \theta = 2\sqrt 2$, we make contact with the integrable Calogero--Moser model, for which extensive literature exists. Even in the more general case, we can make use of the technology of integrable systems, such as the Lax matrix.

\bigskip

Throughout this work we have assumed that the model at the \ac{ir} fixed point is invariant under parity.
This is not a priori necessary and if we relax this assumption, an extra term, scaling as \(\order{Q^{1/4}}\), can appear in the formula for the dimension of the lowest fixed-charge primary.
However, such a term is forbidden for simple algebraic reasons in systems with \(SU(2)\) symmetry, such as the non-linear sigma model used in Section~\refstring{sec:non-linear-sigma-model}.
This seems to match with the experimental observation~\cite{2013PhRvB.88m4411N} that the \(\CP{N-1}\) model flows to a parity-invariant conformal point for \(N = 3 \), while for \(N > 3\) it undergoes a first-order phase transition (which is again second order in the limit \(N \gg 1\)).

\bigskip
An obvious next step is to extend our explicit calculations to $SU(N)$ matrix models with rank $N>3$, which have richer properties than the $SU(3)$ case. For $n>3$, there will be homogeneous solutions with more than one charge which is qualitatively different from the $O(N)$ vector model. The algebraic properties are more intricate than for the case studied here, but our methods are nonetheless applicable.

The other obvious generalization is the study of non-homogeneous solutions, a first example of which has been discussed in~\cite{Hellerman:2017efx}.  Even in the case of the $SU(3)$ matrix model, there are non-homogeneous solutions with more than one fixed charge that can be studied with the methods presented in this paper.

\section*{Acknowledgments}
We would like to thank Luis~Alvarez-Gaume, Antonio~Amariti, Simeon~Hellerman, David~Pirtskhalava and Uwe-Jens~Wiese for enlightening discussions and comments.

The work of  O.L. and S.R. is supported by the Swiss National Science Foundation (\textsc{snf}) under grant number \textsc{pp}00\textsc{p}2\_157571/1.

\appendix
\section{Conventions}\label{app:Conventions}
\addcontentsline{toc}{section}{Appendix}

The Gell--Mann basis for the generators of $A_2$ is given by:
\begin{equation}
  \begin{aligned}
        \lambda_1 &=
        \begin{pmatrix}
        0 & 1 & 0\\
        1 & 0 & 0\\
        0 & 0 & 0
        \end{pmatrix}
        , &
        \lambda_2 &=
        \begin{pmatrix}
        0 & -i & 0\\
        i & 0 & 0\\
        0 & 0 & 0
        \end{pmatrix}
        , &
        \lambda_3 &=
        \begin{pmatrix}
        1 & 0 & 0\\
        0 & -1 & 0\\
        0 & 0 & 0
        \end{pmatrix}
        ,
        \\
        \lambda_4 &=
        \begin{pmatrix}
        0 & 0 & 1\\
        0 & 0 & 0\\
        1 & 0 & 0
        \end{pmatrix}
        , &
        \lambda_5 &=
        \begin{pmatrix}
        0 & 0 & -i\\
        0 & 0 & 0\\
        i & 0 & 0
        \end{pmatrix}
        ,
        \\
        \lambda_6 &=
        \begin{pmatrix}
        0 & 0 & 0\\
        0 & 0 & 1\\
        0 & 1 & 0
        \end{pmatrix}
        , &
        \lambda_7 &=
        \begin{pmatrix}
        0 & 0 & 0\\
        0 & 0 & -i\\
        0 & i & 0
        \end{pmatrix}
        , &
        \lambda_8 &= \frac{1}{\sqrt{3}}
        \begin{pmatrix}
        1 & 0 & 0\\
        0 & 1 & 0\\
        0 & 0 & -2
        \end{pmatrix}
        ,
      \end{aligned}
\end{equation}
normalized as $\Tr \lambda^i \lambda^j = 2 \delta^{ij}$.
The symmetric coefficients $d^{abc}$ defined through
\begin{equation}
  \acomm{\lambda^a}{ \lambda^b} =
  \frac{4}{3} \delta^{ab} + d^{abc} \lambda^c
\end{equation}
are given in the case of $A_2$ algebra by
\begin{equation}
  \begin{aligned}
    d^{118}&= d^{228} = d^{338} = - d^{888} = \frac{2}{\sqrt 3}\\
    d^{448}&=d^{558} = d^{668} = d^{778} = - \frac{1}{\sqrt 3}
    \\
    d^{146}&=d^{157}=-d^{247}=d^{256}=d^{344}=d^{355}=-d^{366} = -d^{377} = 1.
  \end{aligned}
\end{equation}
In total, we have for the product of two Gell--Mann matrices,
\begin{equation}
  \label{app:Product_GellMannMatrices}
  \lambda^a \lambda^b = \frac{2}{3} \delta^{ab} \Id + \frac12\left( d^{abc} + i f^{abc}  \right) \lambda^c .
\end{equation}
Then, it follows for the commutator in these conventions
\begin{equation}
  \left[\lambda^a,\lambda^b\right] = i f^{abc} \lambda^c,
\end{equation}
expressed in terms of the totally antisymmetric structure constants
\begin{equation}
  \begin{aligned}
    f^{123} &= 2 \\
    f^{147}&=-f^{156} = f^{246}=f^{257}=f^{345}=-f^{367} = 1 \\
    f^{458}&=f^{678} = \sqrt{3} .
  \end{aligned}
\end{equation}

\printbibliography

\end{document}